\documentclass{article}
\usepackage{amssymb}

\usepackage{graphicx}
\usepackage{amsmath}

\input{tcilatex}

\begin{document}

\author{V. S. Borisov\thanks{%
E-mail: vyacheslav.borisov (at) gmail.com}}
\title{On dynamics of nonmagnetic accretion disks}
\maketitle

\begin{abstract}
Axisymmetric accretion disks in vicinity of a central compact body are
studied. In the case of non-viscous disk it is proven that all solutions for
the midplane circular velocity are unstable. Hence, the pure hydrodynamic
turbulence\ in accretion disks is possible. It is disproved the well-known
arguments that an inviscid accretion disk must be sub-Keplerian. It is also
demonstrated that the regular asymptotic solutions, often used in
astrophysics, can lead to erroneous conclusions. It is proven that a laminar
viscous disk can be approximated with a great precision by the vortex
motion. Assuming that a turbulent gas tends to flow with minimal losses, we
have shown that a turbulent disk tends to be Keplerian.
\end{abstract}

\section{Introduction\label{Introduction}}

As it was said in \cite{Kurbatov et al. 2014}: ``One of the major challenges
in modern astrophysics is the unexplained turbulence of gas-dynamic
(nonmagnetic) accretion disks. Since they are stable, such disks should not
theoretically be turbulent, but observations show they are. The search for
instabilities that can develop into turbulence is one of the most intriguing
problems in modern astrophysics.'' The question of pure hydrodynamic
turbulence is still under discussion (see, e.g., \cite{Boss 2005}, \cite
{Frank et al. 2002}, \cite{Malanchev et al. 2019}, \cite{Marcus et al. 2015}%
, \cite{Raettig et al. 2013}, \cite{Razdoburdin 2020}, \cite{Razdoburdin and
Zhuravlev 2015} and references therein). The article \cite{Marcus et al.
2015} begins with the questions: ``Can Non-magnetically Coupled,
Non-self-gravitating Keplerian Disks Have Purely Hydrodynamic Instabilities?
Can They Drive Angular Momentum Transport?'' The authors respond: ``After 40
years of intense theoretical and computational research, the answer has been
a qualified -- and unsatisfying -- ``maybe''.'' Let us, however, note that
inviscid accretion disks can have purely hydrodynamic instabilities, as it
has been proven in \cite{Borisov 2013}. In this paper we continue our
investigations.

The paper is devoted to the dynamics of non-magnetic accretion disks. The
input system of equations is the following\footnote{%
Here and in what follows, the standard tensor notation is used (e.g. \cite
{Reddy 2008}, \cite{Schade and Neemann 2018}). In particular, the double
inner product of two second-order tensors, $\mathbf{A}$ and $\mathbf{B}$ ($%
\left[ \mathbf{A}\right] =\left[ A_{ij}\right] $, $\left[ \mathbf{B}\right] =%
\left[ B_{ij}\right] $, $i,j=1,2,3$), is denoted as $\mathbf{A}:\mathbf{B}%
=\sum_{i}\sum_{j}A_{ij}B_{ji}$.} (see \cite{Clarke and Carswell 2007}, \cite
{Frank et al. 2002}, \cite{Goedbloed and Poedts 2004}, \cite{Lai et al. 2010}%
, \cite{Landau and Lifshitz 1987}, \cite{Loitsyanskiy 1978}, \cite{Reddy
2008}, \cite{Sedov 1971}, \cite{Shapiro and Teukolsky 2004}, \cite{Shore
2007}, \cite{Vietri Mario 2008}, \cite{Warsi 1999}).\newline
Conservation of mass:
\begin{equation}
\frac{\partial \rho }{\partial t}+\nabla \cdot \left( \rho \mathbf{v}\right)
=0,  \label{Int10}
\end{equation}
Conservation of momentum:
\begin{equation}
\rho \frac{\partial \mathbf{v}}{\partial t}+\rho \left( \mathbf{v\cdot }%
\nabla \right) \mathbf{v}=-\nabla P+\nabla \cdot \mathbf{\tau }-\rho \nabla
\Phi \mathbf{,}  \label{Int20}
\end{equation}
Conservation of energy\footnote{%
Let us note that the energy equation in the paper \cite{Borisov 2013}\
contains a misprint.}:
\begin{equation}
\frac{\partial E}{\partial t}+\nabla \cdot \left[ \left( E+P\right) \mathbf{v%
}\right] =-\rho \mathbf{v\cdot }\nabla \Phi +\rho S_{E}-\nabla \cdot \mathbf{%
q}_{hc}\mathbf{+}\nabla \mathbf{\cdot \left( \mathbf{\tau \cdot v}\right) ,}
\label{Int30}
\end{equation}
where $\rho $, $P$, $\mathbf{v}$, $\mathbf{\tau }$, $\mathbf{q}_{hc}$, $\Phi
$, and $S_{E}$ denote respectively the density, pressure, velocity, shear
stress tensor, energy flux due to heat conduction, gravitational potential,
and source term, and $E$ $=$ $\rho e_{p}$ $+$ $0.5\rho v^{2}$ denotes the
total energy per unit volume with $e_{p}$ being the internal energy per unit
mass of the fluid and $v$ $=$ $\left| \mathbf{v}\right| $. The shear stress
tensor, $\mathbf{\tau }$, is the sum of two symmetric tensors, $\mathbf{\tau
=\tau }_{v}+\mathbf{\tau }_{t}$, namely, the viscous, $\mathbf{\tau }_{v}$,
and the turbulent, $\mathbf{\tau }_{t}$, stress tensors:
\begin{equation}
\mathbf{\tau }_{v}\approx \mu _{v}\left[ \nabla \mathbf{v}+\left( \nabla
\mathbf{v}\right) ^{\ast }\right] -\frac{2}{3}\left( \mu _{v}\nabla \cdot
\mathbf{v}\right) \mathbf{I},  \label{Int40}
\end{equation}
\begin{equation}
\mathbf{\tau }_{t}\approx \mu _{t}\left[ \nabla \mathbf{v}+\left( \nabla
\mathbf{v}\right) ^{\ast }\right] -\frac{2}{3}\left( \mu _{t}\nabla \cdot
\mathbf{v+}\rho \overline{k}\right) \mathbf{I},  \label{Int50}
\end{equation}
where $\mathbf{I}$\ is the identity tensor, $\left( \mbox{\hspace{2.0mm}}%
\right) ^{\ast }$ denotes a conjugate tensor, $\mu _{v}$ denotes the dynamic
viscosity, $\mu _{t}$, and $\overline{k}$ denote the turbulent viscosity and
kinetic energy of turbulence, respectively (see, e.g., \cite{Anderson et al.
1984}, \cite{Fridman et al 2006}, \cite{Wilcox 1994} and references
therein). For the sake of simplicity, it is believed that the Stokes
hypothesis is valid, i.e. the coefficient of bulk viscosity is negligible.
We will also use the viscosity $\mu =\mu _{v}+\mu _{t}$. Obviously, if the
flow will be laminar, then $\mu _{t}=0$, $\overline{k}=0$ and, hence, $\mu
=\mu _{v}$ will be the dynamic viscosity.

Recall that the dynamic viscosity, $\mu _{v}$, of a gas increases with
absolute temperature, $T$, and, in fact, it is independent of pressure and
density at a given temperature (see, e.g., \cite[p. 186]{Anderson et al.
1984}, \cite[p. 153]{Clarke and Carswell 2007}, \cite[p. 46]{Landau and
Lifshitz 1987}, \cite[p. 635]{Loitsyanskiy 1978}, \cite[pp. 26-28]{White
2006}). Let us cite a couple of widely used approximations for the viscosity
of dilute gases (see, e.g., \cite{Hirschel 2005}, \cite{Loitsyanskiy 1978},
\cite{White 2006}). The power law:
\begin{equation}
\mu _{v}\varpropto T^{n},  \label{Int54}
\end{equation}
where typically $n=0.76$. It is, also, assumed that $n=1$ for the case of
low temperatures. If the temperature is relatively high, then $n\gtrapprox
0.5$, as it is evident from the Sutherland's law:
\begin{equation}
\mu _{v}\varpropto \frac{T^{1.5}}{T+const}.  \label{Int56}
\end{equation}

The turbulent viscosity, $\mu _{t}$, and the turbulence kinetic energy, $%
\overline{k}$, can be estimated accurately by a $\overline{k}$-$\varepsilon $
model (see, e.g., \cite{Archambaul 2002}, \cite{Jin-Lu Yu 2013}, \cite
{Pittard et al. 2009}, \cite{Tominaga and Stathopoulos 2009}) and, hence, it
can be used as the ``principal tool'' for the evaluation. In such a case one
needs to solve two additional partial differential equations (PDEs). For
mildly complex flows, the $\overline{k}$-$\varepsilon $ model can be reduced
to the one-equation model for the kinetic energy of turbulence using Prandtl
and Kolmogorov suggestion (e.g., \cite[p. 230]{Anderson et al. 1984},
\cite[p. 74]{Wilcox 1994}). It is assumed that the turbulent viscosity is
proportional to $\left( \overline{k}\right) ^{0.5}$. In such a case, the
turbulent viscosity, $\mu _{t}$, can be evaluated as the following.
\begin{equation}
\mu _{t}=C_{\overline{k}}\rho \left( \overline{k}\right) ^{0.5}L,\quad C_{%
\overline{k}}=const,  \label{Int60}
\end{equation}
where $L$ denotes the turbulence length scale. The balance equation for the
kinetic energy of turbulence, $\overline{k}$, can be written in the
following form (e.g., \cite{Anderson et al. 1984}, \cite{Wilcox 1994}):
\begin{equation}
\frac{\partial \rho \overline{k}}{\partial t}+\nabla \rho \overline{k}%
\mathbf{v}=\nabla \cdot \left( \mu \nabla \overline{k}\right) +\mathbf{\tau }%
_{t}:\left( \nabla \mathbf{v}\right) -\rho \frac{C_{\mu }\left( \overline{k}%
\right) ^{1.5}}{L},\quad C_{\mu }=const.  \label{Int70}
\end{equation}

Thus, in the case of one-equation model, (\ref{Int60}) and (\ref{Int70}),
the length scale of turbulence is a free parameter in contrast to the
two-equation models (e.g., $\overline{k}$-$\varepsilon $ models), which are
complete and, hence, can be used for computation of the kinetic energy, $%
\overline{k}$, as well as the turbulence length scale, $L$, or equivalent
\cite{Wilcox 1994}. Because of it, most of these models are in widespread
use (e.g., \cite{Archambaul 2002}, \cite{Jin-Lu Yu 2013}, \cite{Pittard et
al. 2009}, \cite{Tominaga and Stathopoulos 2009}). It is interesting to note
that, the so called, algebraic (zero-equation \cite{Wilcox 1994}) models
enjoy widespread use because of their simplicity. For instance, the widely
known $\alpha $-Disc model (e.g., \cite{Armijo 2012}, \cite{Piran 1978},
\cite{Regev 1983}, \cite{Shakura and Sunyaev 1973}) is algebraic, namely,
the turbulent viscosity is evaluated by (\ref{Int60}), where, in fact, the
square root of kinetic energy is assumed to be equal to the sound velocity
and the turbulence length scale is equal to the disk semi-thickness \cite
{Regev 1983}. Interestingly, the $\alpha $-Disc model is unstable (see,
e.g., \cite{Armijo 2012}, \cite{Piran 1978}, \cite{Pringle 1981}).

By and large a perfect\footnote{%
The terms perfect gas and ideal gas are used interchangeably in this paper.}
gas flow will be discussed in this paper and, hence,
\begin{equation}
P=\rho RT,\quad R=const.  \label{Int75}
\end{equation}
We will in general consider axisymmetric gas flows in cylindrical
coordinates, $\left( r,\varphi ,z\right) $. The gravitational potential is
assumed to be as follows
\begin{equation}
\Phi =-G\frac{M}{\sqrt{r^{2}+z^{2}}},\qquad G,M=const.  \label{Int80}
\end{equation}

It will be mainly assumed that the absolute temperature $T$\ is a ``free''
parameter, i.e. $T=T\left( r,z,t\right) $ is a free (or pre-assigned)
function of the coordinates in the region. Such an approach permits us to
avoid using the energy equation. Nevertheless, the energy equation (\ref
{Int30}) will sometimes be replaced by the polytropic relation:
\begin{equation}
P=K\rho ^{\gamma },\quad K,\gamma =const.  \label{Int85}
\end{equation}
If the process is polytropic, then one obtains the following simple formula
\cite[Sec. 5.4.8]{Sedov 1971} for the case of perfect gas:
\begin{equation}
dq^{\left( e\right) }=c_{v}\frac{\gamma -\frac{c_{p}}{c_{v}}}{\gamma -1}dT,
\label{Int90}
\end{equation}
where $dq^{\left( e\right) }$ denotes the external heat flow, $c_{v}$ and $%
c_{p}$ denote the specific heat capacities at constant volume and pressure,
respectively. Let, for definiteness, $dT>0$. In such a case, if $1$ $<$ $%
c_{p}\diagup c_{v}$ $<$ $\gamma $ (or $1$ $<$ $\gamma $ $<$ $c_{p}\diagup
c_{v}$) then $dq^{\left( e\right) }>0$ (or, respectively, $dq^{\left(
e\right) }<0$), i.e. heat is supplied (or, respectively, heat is released)
with an increase in temperature. If $\gamma $ $=$ $c_{p}\diagup c_{v}$, then
$dq^{\left( e\right) }=0$ and, hence, such a polytropic process will be
adiabatic. Thin disk accretion must be highly nonadiabatic, as emphasized in
\cite{Shapiro and Teukolsky 2004}. We will assume in such a case that $%
\gamma $ $\neq $ $c_{p}\diagup c_{v}$.

Let us introduce the following dimensional characteristic quantities: $%
t_{\ast }$, $l_{\ast }$, $\rho _{\ast }$, $v_{\ast }$, $p_{\ast }$, $T_{\ast
}$, $\mu _{\ast }$, and $\overline{k}_{\ast }$ for, respectively, time,
length, density, velocity, pressure, temperature, viscosity, and kinetic
energy of turbulence. The characteristic quantity for sound speed $c_{s\ast
}\equiv \sqrt{p_{\ast }\diagup \rho _{\ast }}$. The following notation will
also be used:
\begin{equation*}
S_{h}=\frac{l_{\ast }}{v_{\ast }t_{\ast }},\ E_{u}=\frac{p_{\ast }}{\rho
_{\ast }v_{\ast }^{2}}=\frac{c_{s\ast }^{2}}{v_{\ast }^{2}},
\end{equation*}
\begin{equation}
F_{r}=\frac{v_{\ast }^{2}l_{\ast }}{GM},\ R_{e}=\frac{\rho _{\ast }v_{\ast
}l_{\ast }}{\mu _{\ast }},\ \vartheta _{\overline{k}e}=\frac{2\overline{k}%
_{\ast }}{3v_{\ast }^{2}},  \label{Int100}
\end{equation}
where $S_{h}$, $E_{u}$, $F_{r}$, and $R_{e}$ denote, respectively, Strouhal,
Euler, Froude, and Reynolds numbers. Throughout of this paper, we assume
that $c_{s\ast }\ll v_{k\ast }\equiv \sqrt{GM\diagup l_{\ast }}$ (see, e.g.,
\cite{Frank et al. 2002}, \cite{Vietri Mario 2008}). Hence, if $v_{\ast
}=v_{k\ast }$, then $E_{u}\ll 1$ and $F_{r}=1$. Note that since $c_{s\ast
}\ll v_{k\ast }$, then $E_{u}F_{r}\ll 1$ even if $v_{\ast }\neq v_{k\ast }$.
In what follows we assume that
\begin{equation}
E_{u}\ll 1,\quad E_{u}\ll \frac{1}{F_{r}}.  \label{Int102}
\end{equation}

It is significant that the Reynolds number, $R_{e}$, is very high (see,
e.g., \cite[p. 170]{Clarke and Carswell 2007}, \cite[p. 70]{Frank et al.
2002}, \cite[p. 165]{Garcia 2011}), i.e.
\begin{equation}
R_{e}\sim 10^{14}.  \label{Int105}
\end{equation}

Let $\varsigma $ denote a variable and let $\varsigma _{\ast }$ denote the
characteristic quantity for $\varsigma $, then the transformation from
dimensional to dimensionless variables can be written in the following form:
\begin{equation}
\varsigma \ \rightarrow \ \varsigma _{\ast }\varsigma .  \label{Int110}
\end{equation}
Notice that we use, as a rule, the characteristic quantity $\mu _{\ast
}\equiv \mu _{v\ast }$, where $\mu _{v\ast }$ denotes the characteristic
quantity for the dynamic viscosity, $\mu _{v}$. In such a case, in view of (%
\ref{Int110}), $\mu $ $\rightarrow $ $\mu _{v\ast }\mu $ $\equiv $ $\mu
_{v\ast }\left( \mu _{v}+\mu _{t}\right) $. However, it can, often, be
convenient to use the characteristic quantity $\mu _{t\ast }$\ in addition
to $\mu _{v\ast }$. Then, the transformation (\ref{Int110}), for the case of
viscosity, can be written as follows:
\begin{equation}
\mu \ \rightarrow \ \mu _{v\ast }\mu \equiv \mu _{v\ast }\left( \mu _{v}+%
\frac{\mu _{t\ast }}{\mu _{v\ast }}\mu _{t}\right) .  \label{Int120}
\end{equation}

For the axisymmetrical flow, we have, in view of (\ref{Int100}) and (\ref
{Int110}), the following non-dimensional system of PDEs.

The continuity equation:
\begin{equation}
S_{h}\frac{\partial \rho }{\partial t}+\frac{1}{r}\frac{\partial \left(
r\rho v_{r}\right) }{\partial r}+\frac{\partial \left( \rho v_{z}\right) }{%
\partial z}=0,  \label{Int130}
\end{equation}

Conservation of momentum:

The $r$-component:
\begin{equation*}
S_{h}\frac{\partial \rho v_{r}}{\partial t}+\frac{1}{r}\frac{\partial }{%
\partial r}r\left( \rho v_{r}^{2}\right) +\frac{\partial }{\partial z}\left(
\rho v_{r}v_{z}\right) -\frac{\rho v_{\varphi }^{2}}{r}=-\frac{\partial }{%
\partial r}\left( E_{u}P\right) -\frac{\rho }{F_{r}}\frac{\partial \Phi }{%
\partial r}+
\end{equation*}
\begin{equation*}
\frac{1}{R_{e}}\left\{ \frac{\partial }{\partial r}\left[ 2\mu \frac{%
\partial v_{r}}{\partial r}-\frac{2}{3}\mu \left( \frac{1}{r}\frac{\partial
\left( rv_{r}\right) }{\partial r}+\frac{\partial v_{z}}{\partial z}\right) %
\right] +\right.
\end{equation*}
\begin{equation}
\left. \frac{\partial }{\partial z}\mu \left( \frac{\partial v_{r}}{\partial
z}+\frac{\partial v_{z}}{\partial r}\right) +\frac{2\mu }{r}\left( \frac{%
\partial v_{r}}{\partial r}\mathbf{-}\frac{v_{r}}{r}\right) \right\}
-\vartheta _{\overline{k}e}\frac{\partial }{\partial r}\rho \overline{k},
\label{Int140}
\end{equation}

The $\varphi $-component:
\begin{equation*}
S_{h}\frac{\partial \rho v_{\varphi }}{\partial t}+\frac{1}{r}\frac{\partial
}{\partial r}r\left( \rho v_{\varphi }v_{r}\right) +\frac{\partial }{%
\partial z}\left( \rho v_{\varphi }v_{z}\right) +\frac{\rho v_{\varphi }v_{r}%
}{r}=
\end{equation*}
\begin{equation}
\frac{1}{R_{e}}\left\{ \frac{\partial }{\partial r}\left[ \mu r\frac{%
\partial }{\partial r}\left( \frac{v_{\varphi }}{r}\right) \right] +\frac{%
\partial }{\partial z}\left( \mu \frac{\partial v_{\varphi }}{\partial z}%
\right) +2\mu \frac{\partial }{\partial r}\left( \frac{v_{\varphi }}{r}%
\right) \right\} ,  \label{Int150}
\end{equation}

The $z$-component:
\begin{equation*}
S_{h}\frac{\partial \rho v_{z}}{\partial t}+\frac{1}{r}\frac{\partial }{%
\partial r}r\left( \rho v_{z}v_{r}\right) +\frac{\partial }{\partial z}%
\left( \rho v_{z}^{2}\right) =-\frac{\partial }{\partial z}\left(
E_{u}P\right) -\frac{\rho }{F_{r}}\frac{\partial \Phi }{\partial z}+
\end{equation*}
\begin{equation*}
\frac{1}{R_{e}}\left\{ \frac{\partial }{\partial r}\left[ \mu \left( \frac{%
\partial v_{r}}{\partial z}+\frac{\partial v_{z}}{\partial r}\right) \right]
+\frac{\partial }{\partial z}\left[ 2\mu \frac{\partial v_{z}}{\partial z}-%
\frac{2}{3}\mu \left( \frac{1}{r}\frac{\partial \left( rv_{r}\right) }{%
\partial r}+\frac{\partial v_{z}}{\partial z}\right) \right] +\right.
\end{equation*}
\begin{equation}
\left. \frac{\mu }{r}\left( \frac{\partial v_{r}}{\partial z}+\frac{\partial
v_{z}}{\partial r}\right) \right\} -\vartheta _{\overline{k}e}\frac{\partial
}{\partial z}\rho \overline{k},  \label{Int160}
\end{equation}
where
\begin{equation}
\Phi =-\frac{1}{\sqrt{r^{2}+z^{2}}}\ .  \label{Int165}
\end{equation}
For definiteness sake, it will be assumed that $v_{\varphi }\geq 0$. The
following Taylor series of $\Phi $ about $z=0$\ will also be used.
\begin{equation}
\Phi \equiv -\frac{1}{\sqrt{r^{2}+z^{2}}}=\Phi _{0}+\Phi _{2}z^{2}+\ldots =-%
\frac{1}{r}+\frac{1}{2r^{3}}z^{2}+\ldots ,\quad \left| z\right| <r.
\label{Int170}
\end{equation}
We will take $p_{\ast }=\rho _{\ast }RT_{\ast }$ and, hence, we obtain from (%
\ref{Int75}) that
\begin{equation}
P=\rho T.  \label{Int175}
\end{equation}

If the process is polytropic, then, instead of (\ref{Int85}), we have the
following non-dimensional relation.

\begin{equation}
P=\kappa \rho ^{\gamma },\quad \kappa =\frac{K\rho _{\ast }^{\gamma }}{%
p_{\ast }}=const  \label{Int176}
\end{equation}

We will, in general, consider the outer regions (see, e.g. , \cite{Clarke
and Carswell 2007}, \cite{Garcia 2011}, \cite{Hartmann 2009}, \cite{Regev
1983}) of accretion disks, i.e. $r\geq r_{0}$ (say), where $r_{0}\gg
r_{s}+b_{s}$, $r_{s}$ denotes the non-dimensional radius of star, $b_{s}$
denotes the non-dimensional radial extent of boundary layer (e.g.,
\cite[Sec. 6.2]{Frank et al. 2002}). For the sake of simplicity, we take the
characteristic quantity, $l_{\ast }$, for length such that $r_{0}=1$ and,
hence, $r_{s}+b_{s}\ll 1$. Thus, the gas flows will be considered in the
region $\frak{R}_{a}$=$\left\{ \left( r,z\right) :\text{ }r\in \left[
1,\infty \right) \text{, }z\in \left[ 0,\infty \right) \right\} $, due to
the axisymmetry of the problem.

Let $\varsigma $ denote a dependent variable. It will be used throughout
this paper that
\begin{equation}
\left. \varsigma \right| _{r\rightarrow \infty }=\left. \varsigma \right|
_{z\rightarrow \infty },\quad \hat{\varsigma}\equiv \left. \varsigma \right|
_{r\rightarrow 1}.  \label{Int177}
\end{equation}

Following the widespread view on thin accretion disk dynamics (see, e.g.,
\cite{Clarke and Carswell 2007}, \cite{Frank et al. 2002}, \cite{Garcia 2011}%
, \cite{Hartmann 2009}, \cite{Shapiro and Teukolsky 2004}, \cite{Vietri
Mario 2008}) we will, mainly, consider the case when the gas possesses a
small inward velocity, i.e. the radial drift velocity is highly subsonic:
\begin{equation}
\left| v_{r}\right| \ll c_{s},  \label{Int180}
\end{equation}
where $c_{s}$ denotes the sound speed. We will, as usually, assume that a
physically correct model must accurately simulate Keplerian disks since the
majority of observed disks are in Keplerian (or sub-Keplerian) rotation
around their central accreting objects (see, e.g., \cite{Harsono et al. 2013}%
, \cite{Hure et al. 2011}, \cite{Murillo et al. 2013}, \cite{Simon et al.
2001}, \cite{Takakuwa et al. 2012}, \cite{Yen et al. 2014} and references
therein).

Since we consider symmetric disks, it is, in general, assumed that $0\leq
z\leq H$, where $H$ denotes the disk semi-thickness (e.g. \cite{Duric 2004},
\cite{Shapiro and Teukolsky 2004}). The temperature and density at the disk
surface, i.e.
\begin{equation}
T_{H}\equiv \left. T\left( r,z,t\right) \right| _{z=\pm H},\ \rho _{H}\equiv
\left. \rho \left( r,z,t\right) \right| _{z=\pm H},\quad T_{H}\geq 0,\ \rho
_{H}\geq 0,  \label{Int183}
\end{equation}
are pre-assigned functions of $r$ and $t$. In the following we will
sometimes assume, for the sake of convenience, that $T_{H}\approx 0$ and $%
\rho _{H}\approx 0$. It is pertinent to note that
\begin{equation}
\rho =\rho \left( r,z,t\right) >0,\quad \forall z:\ \left| z\right| <H.
\label{Int185}
\end{equation}

It is assumed that the dependent variables are continuous functions of $z$
at $z=0$ and, in this case, the values $\rho $, $P$, $T$, $v_{r}$, $%
v_{\varphi }$, are even functions of $z$, whereas $v_{z}$ is an odd one. We
will also use the following asymptotic expansions in the limit $z\rightarrow
0$:
\begin{equation*}
\rho \sim \rho ^{\circ }+\rho ^{\prime \prime \circ }z^{2}+\ldots ,\quad
T\sim T^{\circ }+T^{\prime \prime \circ }z^{2}+\ldots ,
\end{equation*}
\begin{equation}
P\sim P^{\circ }+P^{\prime \prime \circ }z^{2}+\ldots ,  \label{Int190}
\end{equation}
\begin{equation*}
v_{r}\sim v_{r}^{\circ }+v_{r}^{\prime \prime \circ }z^{2}+\ldots ,\
v_{\varphi }\sim v_{\varphi }^{\circ }+v_{\varphi }^{\prime \prime \circ
}z^{2}+\ldots ,
\end{equation*}
\begin{equation}
v_{z}\sim v_{z}^{\prime \circ }z+v_{z}^{\prime \prime \prime \circ
}z^{3}+\ldots ,  \label{Int200}
\end{equation}
where the symmetry was taken into account. As usually (e.g., \cite{Jones
1997}), the series in (\ref{Int190})-(\ref{Int200}) may converge or diverge,
but their partial sums are good approximations to the dependent variables
for small enough $z$.

We will, in general, consider geometrically thin accretion disks and, hence,
it assumed (see, e.g., \cite[p. 157]{Duric 2004}, \cite[p. 87, p. 129]{Frank
et al. 2002}, \cite[p. 65]{Fridman et al 2006}, \cite[p. 432]{Shapiro and
Teukolsky 2004}, \cite[p. 304]{Vietri Mario 2008}) that the semi-thickness, $%
H$, everywhere satisfies
\begin{equation}
\frac{H}{r}\ll 1,\quad \left| \frac{\partial H}{\partial r}\right| \ll 1.
\label{Int205}
\end{equation}

Sometimes, for the sake of convenience, we will use the set of hyper-real
numbers ($^{\ast }\mathbb{R}$), which contains the set of real numbers ($%
\mathbb{R}$), the set of infinitesimals (hyper-small numbers), and the set
of infinite (hyper-large) numbers (see, e.g., \cite{Chaudhuri 2016}, \cite
{Jones 1997}, \cite{Keisler 2012}, \cite{Ponstein 2001}). Let us recall the
terminology. If $\iota $ (iota) is such that $\left| \iota \right| $ $<a$
for every real $a>0$, then $\iota $ is called infinitesimal or hyper-small
number. There is only one real number that is infinitesimal and that is $0$.
If $\iota $ will be hyper-small but non-zero, then $\omega =1\diagup \iota $
will be hyper-large, that is, $\left| \omega \right| $ will be greater than
any real number. The hyper-large numbers must not be confused with infinity (%
$\infty $), which is not a number at all. We will use the following
notation: $a$ $\cong $ $b$ means that a number $a$ is infinitely close to a
number $b$, i.e. their difference $a-b$ is infinitesimal, and $a$ $\ncong $ $%
b$ means that a number $a$ is not infinitely close to another one $b$. The
notation $a$ $\simeq $ $b$ means that $a$ $\cong $ $b$ but $a$ $\neq $ $b$.
Let $a$ be a finite hyper-real number. The real number which is infinitely
close to $a$ is called the standard part of $a$ and denoted by $st\left(
a\right) $, and, hence, $st\left( a\right) $ $\cong $ $a$. Obviously, $%
st\left( a\right) $ $=$ $a$ if and only if (shortened iff) $a\in \mathbb{R}$%
. Let $f:$ $^{\ast }\mathbb{R\rightarrow }^{\ast }\mathbb{R}$. The
hyper-real number $L$ is the limit\footnote{%
It can be written in the following form \cite{Ponstein 2001}: $L=\underset{%
x\rightarrow a}{\lim }f(x)$ if $\forall \varepsilon $ $\in $ $^{\ast }%
\mathbb{R}$, $\varepsilon >0$, $\exists \delta $ $\in $ $^{\ast }\mathbb{R}$%
, $\delta >0$ $:$ $\forall x$ $\in $ $^{\ast }\mathbb{R}$, $0<\left|
x-a\right| <\delta $ $\Rightarrow $ $\left| f(x)-L\right| <\varepsilon $} of
$f(x)$ as $x$ $\in $ $^{\ast }\mathbb{R}$ approaches $a$ $\in $ $^{\ast }%
\mathbb{R}$ (i.e. $\underset{x\rightarrow a}{\lim }f(x)=L$) if whenever $x$ $%
\simeq $ $a$, $f(x)$ $\cong $ $L$. The hyper-real number $f^{\prime }\left(
x\right) $ will be the derivative of $f:$ $^{\ast }\mathbb{R\rightarrow }%
^{\ast }\mathbb{R}$ at $x$ $\in $ $^{\ast }\mathbb{R}$ iff
\begin{equation}
f^{\prime }\left( x\right) =\underset{\Delta x\rightarrow 0}{\lim }\frac{%
f\left( x+\Delta x\right) -f\left( x\right) }{\Delta x},\quad \Delta x\simeq
0.  \label{Int210}
\end{equation}
The real number $F^{\prime }\left( x\right) $ will be the S-derivative of
the $f(x)$ at $x$ $\in $ $^{\ast }\mathbb{R}$ iff
\begin{equation}
F^{\prime }\left( x\right) =st\left( \frac{f\left( x+\Delta x\right)
-f\left( x\right) }{\Delta x}\right) ,\quad \forall \,\Delta x\simeq 0.
\label{Int220}
\end{equation}
Notice that $f^{\prime }\left( x\right) -F^{\prime }\left( x\right) $ $\cong
$ $0$, i.e. $F^{\prime }\left( x\right) =st\left[ f^{\prime }\left( x\right)
\right] $.

In connection with the aforesaid, the assumption (\ref{Int185}) should be
adjusted. To be precise, we should write that $\rho $ $\ncong $ $0$, namely $%
st\left( \rho \right) >0$, if $\left| z\right| <H$. It is inconvenient to
make constant reference to hyper-real numbers in this paper. So, a phrase
such as the above one will often be abbreviated to (\ref{Int185}) with any
definitions understood implicitly. It should induce no difficulty so long as
any more precise definitions are mentioned explicitly when that is vital.

We will, in general, use the same symbol, $\partial $, for partial as well
as ordinary derivatives of a function even if the function depends on one
variable only. The symbol $d$ will be used for the total (full) derivative
of a function.

\section{Laminar Flow}

Only laminar flows are considered in this section and, hence, the turbulent
viscosity $\mu _{t}=0$, and the kinetic energy of turbulence $\overline{k}=0$%
.

\subsection{Inviscid Flow}

We consider the outer regions of accretion disks, where the damping due to
molecular viscosity is very small (e.g., \cite[p. 70]{Boss 2005}, \cite[p.
70]{Frank et al. 2002}, \cite[p. 143]{Hartmann 2009}) and, hence, we expect
that the friction forces are negligible, i.e. the Reynolds numbers are very
high (say, $R_{e}\sim 10^{14}$, e.g., \cite[p. 170]{Clarke and Carswell 2007}%
, \cite[p. 70]{Frank et al. 2002}, \cite[p. 165]{Garcia 2011}). Therefore,
the solutions of the Navier-Stokes equations will be such that the outer
flow obeys the laws of inviscid flow \cite{Schlichting 1968}. In this
connection, only inviscid flows are considered in this sub-section and,
hence, the dynamic viscosity $\mu _{v}=0$. The steady-state version of the
PDE system (\ref{Int130})-(\ref{Int160}) is thus reduced to the following
one.
\begin{equation}
\frac{1}{r}\frac{\partial \left( r\rho v_{r}\right) }{\partial r}+\frac{%
\partial \left( \rho v_{z}\right) }{\partial z}=0,  \label{InvF10}
\end{equation}
\begin{equation*}
\frac{1}{r}\frac{\partial }{\partial r}r\left( \rho v_{r}^{2}\right) +\frac{%
\partial }{\partial z}\left( \rho v_{r}v_{z}\right) \equiv
\end{equation*}
\begin{equation}
\rho v_{r}\frac{\partial v_{r}}{\partial r}+\rho v_{z}\frac{\partial v_{r}}{%
\partial z}=\frac{\rho v_{\varphi }^{2}}{r}-E_{u}\frac{\partial P}{\partial r%
}-\frac{\rho }{F_{r}}\frac{\partial \Phi }{\partial r},  \label{InvF20}
\end{equation}
\begin{equation*}
\frac{1}{r}\frac{\partial }{\partial r}r\left( \rho v_{\varphi }v_{r}\right)
+\frac{\partial }{\partial z}\left( \rho v_{\varphi }v_{z}\right) +\rho
\frac{v_{\varphi }v_{r}}{r}\equiv
\end{equation*}
\begin{equation}
\rho v_{r}\frac{\partial v_{\varphi }}{\partial r}+\rho v_{z}\frac{\partial
v_{\varphi }}{\partial z}+\rho \frac{v_{\varphi }v_{r}}{r}=0,  \label{InvF30}
\end{equation}
\begin{equation}
\rho v_{r}\frac{\partial v_{z}}{\partial r}+\rho v_{z}\frac{\partial v_{z}}{%
\partial z}=-E_{u}\frac{\partial P}{\partial z}-\frac{\rho }{F_{r}}\frac{%
\partial \Phi }{\partial z},  \label{InvF40}
\end{equation}
where $\Phi $\ is defined by Eq. (\ref{Int165}), i.e. $\Phi =-\left(
r^{2}+z^{2}\right) ^{-0.5}$. In what follows the perfect gas equation of
state, (\ref{Int175}), will, mainly, be used, i.e. $P=\rho T$.

Taking some liberties with the notation (\ref{Int177}), we write
\begin{equation}
\left. \left( v_{r},v_{\varphi },v_{z}\right) \right| _{r\rightarrow \infty
}=\left. \left( v_{r},v_{\varphi },v_{z}\right) \right| _{z\rightarrow
\infty }\rightarrow 0,  \label{InvF50}
\end{equation}
\begin{equation}
\left. \left( \rho ,P,T\right) \right| _{r\rightarrow \infty }=\left. \left(
\rho ,P,T\right) \right| _{z\rightarrow \infty }\rightarrow \left( \rho
_{\infty },P_{\infty },T_{\infty }\right) =\mathbf{const}.\,  \label{InvF60}
\end{equation}
The following notations will also be used.
\begin{equation}
\rho ^{\circ }\equiv \left. \rho \right| _{z=0}>0,\quad P^{\circ }\equiv
\left. P\right| _{z=0},\quad T^{\circ }\equiv \left. T\right| _{z=0},
\label{InvF70}
\end{equation}
\begin{equation}
v_{r}^{\circ }\equiv \left. v_{r}\right| _{z=0},\quad v_{\varphi }^{\circ
}\equiv \left. v_{\varphi }\right| _{z=0},\quad v_{z}^{\prime \circ }\equiv
\left. \frac{\partial v_{z}}{\partial z}\right| _{z=0}.  \label{InvF75}
\end{equation}

\subsubsection{Circular velocity \label{Circular velocity}}

Since the values $\rho $, $P$, $v_{r}$, and $v_{\varphi }$ are even
functions of $z$, whereas $v_{z}$ is an odd one, the PDE system (\ref{InvF10}%
)-(\ref{InvF30}) at the midplane ($z=0$) becomes the following ODE (ordinary
differential equation) system in the midplane variables.
\begin{equation}
\frac{1}{r}\frac{\partial \left( r\rho ^{\circ }v_{r}^{\circ }\right) }{%
\partial r}+\rho ^{\circ }v_{z}^{\prime \circ }=0,  \label{InvF80}
\end{equation}

\begin{equation}
\rho ^{\circ }v_{r}^{\circ }\frac{\partial v_{r}^{\circ }}{\partial r}-\frac{%
\rho ^{\circ }\left( v_{\varphi }^{\circ }\right) ^{2}}{r}=-E_{u}\frac{%
\partial P^{\circ }}{\partial r}-\frac{1}{F_{r}}\frac{\rho ^{\circ }}{r^{2}},
\label{InvF90}
\end{equation}
\begin{equation}
\rho ^{\circ }v_{r}^{\circ }\frac{\partial rv_{\varphi }^{\circ }}{r\partial
r}=0.  \label{InvF100}
\end{equation}
Eq. (\ref{InvF40}) is satisfied identically at the midplane. Assuming that $%
v_{r}^{\circ }\neq 0$ we find the following exact solution for the midplane
value, $\left. v_{\varphi }\right| _{z=0}$, of circular velocity.
\begin{equation}
\left. v_{\varphi }\right| _{z=0}\equiv v_{\varphi }^{\circ }=\frac{%
C_{\varphi }^{\circ }}{r},\quad C_{\varphi }^{\circ }=const.  \label{InvF110}
\end{equation}
For the sake of convenience, let us represent $C_{\varphi }^{\circ }$ as a
function of the Froude number, $F_{r}$. Let the point $r=r_{m}=const$ ($%
0\leq r_{m}\leq 1$)\ be the only point where the vortex motion, (\ref
{InvF110}), will also be Keplerian:
\begin{equation}
v_{\varphi k}=\frac{1}{\sqrt{rF_{r}}}.  \label{InvF114}
\end{equation}
Then $C_{\varphi }^{\circ }\diagup r_{m}=1\diagup \sqrt{r_{m}F_{r}}$ $%
\Rightarrow $ $C_{\varphi }^{\circ }=\sqrt{r_{m}\diagup F_{r}}$ and, hence,
\begin{equation}
v_{\varphi }^{\circ }=\frac{\sqrt{r_{m}}}{r\sqrt{F_{r}}},\quad r\geq
r_{m},\quad 0\leq r_{m}\leq 1.  \label{InvF116}
\end{equation}

Note that the exact solution, (\ref{InvF110}), was found under the only
assumption of non-zero inward drift velocity, i.e. $v_{r}^{\circ }\equiv
\left. v_{r}\right| _{z=0}\neq 0$. Thus, in the case of inviscid flow with $%
v_{r}^{\circ }\neq 0$ we have the vortex\footnote{%
An axisymmetric flow with $v_{\varphi }\varpropto r^{-1}$ will be called as
vortex. In particular, vortex-sinks and vortex-sources are lumped together
as vortices.} as the only solution for the midplane circular velocity, $%
v_{\varphi }^{\circ }$. If, however, $v_{r}^{\circ }=0$, then $v_{\varphi
}^{\circ }$ can be sub-Keplerian or even highly non-Keplerian.

For the sake of simplicity, it was considered the flow at the midplane. One
can readily see that the same results are valid in a more general case.
Actually, let $z=h\left( r\right) $ denote a streamline, namely a line that
is tangent to the meridional velocity vector, ($v_{r},v_{z}$). The kinematic
condition (see, e.g., \cite[p. 165]{Sedov 1971}, \cite[p. 50]{White 2006})
at the streamline will be the following.
\begin{equation}
v_{z}^{h}=v_{r}^{h}\frac{\partial h}{\partial r},\quad v_{z}^{h}\equiv
\left. v_{z}\right| _{z=h\left( r\right) },\,v_{r}^{h}\equiv \left.
v_{r}\right| _{z=h\left( r\right) }.  \label{InvF120}
\end{equation}
Let $\varsigma =\varsigma \left( r,z\right) $ denote a dependent variable.
Let the point $\left( r,z\right) $ be at the streamline $z=h\left( r\right) $%
, then the total (full) derivative of $\varsigma ^{h}\equiv \left. \varsigma
\right| _{z=h\left( r\right) }$ with respect to $r$ is the following.
\begin{equation}
\frac{d\varsigma ^{h}}{dr}=\frac{\partial \varsigma ^{h}}{\partial r}+\frac{%
\partial \varsigma ^{h}}{\partial z}\frac{\partial h}{\partial r}.
\label{InvF130}
\end{equation}
In view of (\ref{InvF120}) and (\ref{InvF130}), we obtain from (\ref{InvF30}%
):
\begin{equation}
v_{r}^{h}\frac{drv_{\varphi }^{h}}{dr}=0,\quad v_{\varphi }^{h}\equiv \left.
v_{\varphi }\right| _{z=h\left( r\right) }.  \label{InvF140}
\end{equation}
If $v_{r}^{h}\equiv \left. v_{r}\right| _{z=h\left( r\right) }\neq 0$, then,
in view of (\ref{InvF140}), we obtain
\begin{equation}
v_{\varphi v}^{h}\equiv v_{\varphi }^{h}=\frac{C_{\varphi }^{h}}{r},\quad
\left. C_{\varphi }^{h}\right| _{z=h\left( r\right) }=const.  \label{InvF150}
\end{equation}
Thus, in the case of inviscid flow with $v_{r}^{h}\neq 0$ we have the vortex
(\ref{InvF150}) as the only solution for the circular velocity at the
streamline, $z=h\left( r\right) $. We emphasize that the exact solution, (%
\ref{InvF150}), was found without any assumptions about the equation of
state, the temperature distribution in the disk, and the gravitational field.

It is significant that the circular velocity may differ from the vortex only
if the radial velocity is equal to zero. Let us consider such a possibility,
namely, we will find some of the circular velocities, provided $v_{r}\equiv
0 $ in the region $\frak{R}_{a}\ $(=$\left\{ \left( r,z\right) :\text{ }r\in %
\left[ 1,\infty \right) \text{, }z\in \left[ 0,\infty \right) \right\} $).
In such a case we immediately obtain from Eq. (\ref{InvF10}) that $%
v_{z}\equiv 0$, since $\left. v_{z}\right| _{z=0}=0$. Hence, instead of (\ref
{InvF10})-(\ref{InvF40}), we shall have the following PDE system.
\begin{equation}
\frac{\rho v_{\varphi }^{2}}{r}=E_{u}\frac{\partial P}{\partial r}+\frac{%
\rho }{F_{r}}\frac{\partial \Phi }{\partial r},  \label{InvF160}
\end{equation}
\begin{equation}
E_{u}\frac{\partial P}{\partial z}+\frac{\rho }{F_{r}}\frac{\partial \Phi }{%
\partial z}=0,\quad \Phi =-\left( r^{2}+z^{2}\right) ^{-0.5}.
\label{InvF170}
\end{equation}

Let the circular velocity, $v_{\varphi }=v_{\varphi }\left( r,z\right) $, be
a continuous function of $z$ at the midplane $\frak{R}_{a}^{\circ }$ $\equiv
$ $\left\{ \left( r,z\right) :\right. $ $r\in \left[ 1,\infty \right) $, $%
\left. z=0\right\} $. In such a case, $v_{\varphi }\equiv 0$ in the region$%
\frak{\ R}_{a}$ = $\left\{ \left( r,z\right) :\right. $ $r\in \left[
1,\infty \right) $, $\left. z\in \left[ 0,\infty \right) \right\} $ iff the
gas pressure distribution is spherically symmetric, i.e.
\begin{equation}
P=P\left( R\right) ,\quad R=\sqrt{r^{2}+z^{2}}.  \label{InvF175}
\end{equation}
Actually, let (\ref{InvF175}) be valid. Then, in view of (\ref{InvF175}), we
obtain from (\ref{InvF170}) that
\begin{equation}
\frac{\partial P}{\partial R}=-\frac{\rho }{E_{u}F_{r}R^{2}},\quad z>0.
\label{InvF177}
\end{equation}
By virtue of (\ref{InvF177}), we find from (\ref{InvF160}) that $v_{\varphi
}=0$ for all $z>0$ and, hence, in view of the continuity assumption, the
circular velocity $v_{\varphi }\equiv 0$ in the region $\frak{R}_{a}$. Now,
let $v_{\varphi }\equiv 0$ in the region $\frak{R}_{a}$. In such a case we
find from (\ref{InvF160})-(\ref{InvF170}) that
\begin{equation}
z\frac{\partial P}{\partial r}-r\frac{\partial P}{\partial z}=0.
\label{InvF179}
\end{equation}
Then we obtain from (\ref{InvF179}) that (\ref{InvF175}) is valid, i.e. the
gas pressure distribution is spherically symmetric in $\frak{R}_{a}$. Let us
also note that the proof was done without any assumptions about the equation
of state and the temperature distribution in the disk. We emphasize that the
statement, i.e. $v_{\varphi }\equiv 0$ in $\frak{R}_{a}$ $%
\Longleftrightarrow $ $P=P\left( R\right) $ in $\frak{R}_{a}$, has been
proven for the whole region $\frak{R}_{a}$. Let $\frak{R}_{b}$ be a
subregion of $\frak{R}_{a}\ $(i.e. $\frak{R}_{b}\subset \frak{R}_{a}$), then
one can readily see that the similar statement, i.e. $v_{\varphi }\equiv 0$
in $\frak{R}_{b}$ $\Longleftrightarrow $ $P=P\left( R\right) $ in $\frak{R}%
_{b}$, will also be valid. One can also readily see from (\ref{InvF160})-(%
\ref{InvF170}) that $\rho =\rho \left( R\right) $ provided (\ref{InvF175})
and, hence, the temperature distribution of an ideal gas will be spherically
symmetric in $\frak{R}_{a}$ (or in $\frak{R}_{b}$).

Let us now assume that the gas is ideal, (\ref{Int175}), and the temperature
$T=T\left( r,z\right) \geq T_{\infty }>0$ is a pre-assigned function of the
coordinates in the region $\frak{R}_{a}$ = $\left\{ \left( r,z\right)
:\right. $ $r\in \left[ 1,\infty \right) $, $\left. z\in \left[ 0,\infty
\right) \right\} $. In such a case, in view of (\ref{Int175}) and (\ref
{InvF60}), we obtain the following exact solution to Eq. (\ref{InvF170}):
\begin{equation}
P=P^{\circ }\exp \left[ -\int\limits_{0}^{z}\frac{\xi }{E_{u}F_{r}T\left(
r,\xi \right) \left( r^{2}+\xi ^{2}\right) ^{1.5}}d\xi \right] ,\quad
P^{\circ }=P^{\circ }\left( r\right) .  \label{IF50}
\end{equation}
It is a simple matter to evaluate $P^{\circ }\left( r\right) $ for each $r$
by virtue of the boundary conditions (\ref{InvF60}). Actually, in view of (%
\ref{IF50}) and (\ref{InvF60}), we have
\begin{equation}
P_{\infty }=P^{\circ }\left( r\right) \Psi _{\infty }\left( r\right) ,\ \Psi
_{\infty }\left( r\right) \equiv \exp \left[ -\int\limits_{0}^{\infty }\frac{%
\xi }{E_{u}F_{r}T\left( r,\xi \right) \left( r^{2}+\xi ^{2}\right) ^{1.5}}%
d\xi \right] .  \label{IF60}
\end{equation}
It is significant that $P_{\infty }\neq 0$ in (\ref{IF60}). Otherwise, as we
can see from (\ref{IF60}), the value $P^{\circ }$ ($=P^{\circ }\left(
r\right) $) may be assigned arbitrarily provided that $\Psi _{\infty }\left(
r\right) =0$. In such a case the problem (\ref{InvF160})-(\ref{InvF170}) is
reduced to the only equation (\ref{InvF160}), where the two functions,
namely $v_{\varphi }\left( r,z\right) $ and $P^{\circ }\left( r\right) $,
are the unknown functions, i.e. the problem is not well-posed in the sense
of Hadamard (e.g., \cite{Anderson et al. 1984}, \cite{Tikhonov and Arsenin
1977}). With this in mind, under $P_{\infty }=0$, we assume that $\Psi
_{\infty }\left( r\right) \neq 0$ for all $r\in \left[ 1,\infty \right) $.
Then $P^{\circ }\left( r\right) \equiv 0$ and, hence, in view of (\ref{IF50}%
), $P\left( r,z\right) \equiv 0$ in the region $\frak{R}_{a}$. Consequently $%
\rho \left( r,z\right) \equiv 0$ in $\frak{R}_{a}$, as the gas is ideal and,
hence, there is no any flow in the whole region, $\frak{R}_{a}$ = $\left\{
\left( r,z\right) :\right. $ $r\in \left[ 1,\infty \right) $, $\left. z\in %
\left[ 0,\infty \right) \right\} $, for lack of gas. Thus, assuming that $%
P_{\infty }\neq 0$ in (\ref{IF60}), we rewrite the exact solution to Eq. (%
\ref{InvF170}) in the following form:
\begin{equation}
P=P_{\infty }\exp \left\{ \int\limits_{z}^{\infty }\frac{\xi }{%
E_{u}F_{r}T\left( r,\xi \right) \left( r^{2}+\xi ^{2}\right) ^{1.5}}d\xi
\right\} ,\quad P_{\infty }=const>0.  \label{IF100}
\end{equation}
Let us note that (\ref{IF100}) leads to (\ref{InvF175}) if the gas
temperature distribution is spherically symmetric in $\frak{R}_{a}$. In view
of (\ref{IF100}), the Eq. (\ref{InvF160}) gives the exact solution for the
circular velocity:
\begin{equation}
v_{\varphi }^{2}=r\frac{T\left( r,z\right) }{F_{r}}\frac{\partial }{\partial
r}\int\limits_{z}^{\infty }\frac{\xi }{T\left( r,\xi \right) \left(
r^{2}+\xi ^{2}\right) ^{1.5}}d\xi +\frac{r^{2}}{F_{r}\left(
r^{2}+z^{2}\right) ^{1.5}}.  \label{IF110}
\end{equation}

For the sake of illustration, let us, first, consider an isothermal flow in
the region $\frak{R}_{a}$. Since $T\equiv T_{\infty }=const>0$, we find, by
virtue of (\ref{IF100}), that
\begin{equation}
P=P_{\infty }\exp \left( \frac{1}{E_{u}F_{r}T_{\infty }R}\right) ,\quad R=%
\sqrt{r^{2}+z^{2}}.  \label{IF120}
\end{equation}
Thus, it is apparent that $v_{\varphi }\equiv 0$ in $\frak{R}_{a}$, since (%
\ref{InvF175}) is valid (cf. \cite{Kluzniak and Kita 2000}). The same
result, i.e. $v_{\varphi }\equiv 0$ in $\frak{R}_{a}$ given that $T\equiv
const$, can be obtained directly from (\ref{IF110}).

We make now use of the well-known power-law model for the temperature
distribution (see, e.g., \cite{Armijo 2012}, \cite{Armitage 2010}, \cite
{Beskin et al 2002}, \cite{Frank et al. 2002}, \cite{Garcia 2011}, \cite
{Hartmann 2009}, \cite{Jimenez-Vicente 2014}, \cite{Min-Kai Lin and Youdin
2015}, \cite{Nelson et al 2013}, \cite{Shapiro and Teukolsky 2004}, \cite
{Shore 2007}, \cite{Shtemler 2009}, \cite{Vietri Mario 2008}). Let us
consider the following generalization of the model:
\begin{equation}
T=\frac{\hat{T}-T_{\infty }}{r^{\tilde{\alpha}}\sqrt{r^{2}+z^{2}}}+T_{\infty
},\quad \hat{T},\,T_{\infty }=const,\ \hat{T}>T_{\infty },\ \tilde{\alpha}%
=const\geq 0.  \label{IF130}
\end{equation}
We obtain from (\ref{IF100})-(\ref{IF110}), in view of (\ref{IF130}), that
\begin{equation}
P=P_{\infty }\left[ 1+\frac{\hat{T}-T_{\infty }}{T_{\infty }r^{\tilde{\alpha}%
}\sqrt{r^{2}+z^{2}}}\right] ^{\beta _{p}},\quad \beta _{p}=\frac{r^{\tilde{%
\alpha}}}{E_{u}F_{r}\left( \hat{T}-T_{\infty }\right) };  \label{IF140}
\end{equation}
\begin{equation}
v_{\varphi }^{2}=\frac{\tilde{\alpha}}{F_{r}}\left( \frac{r^{\tilde{\alpha}}T%
}{\hat{T}-T_{\infty }}\ln \frac{T}{T_{\infty }}-\frac{1}{\sqrt{r^{2}+z^{2}}}%
\right) ,\quad T=\frac{\hat{T}-T_{\infty }}{r^{\tilde{\alpha}}\sqrt{%
r^{2}+z^{2}}}+T_{\infty }.  \label{IF150}
\end{equation}
Taking the midplane circular velocity to be no more than the Keplerian
velocity, i.e. $\left. v_{\varphi }^{2}\right| _{z=0}\leq v_{\varphi
k}^{2}\equiv \left( rF_{r}\right) ^{-1}$ for all $r\in \left[ 1,\infty
\right) $, we obtain an upper bound for the parameter $\tilde{\alpha}$ in (%
\ref{IF130})-(\ref{IF150}). The midplane circular velocity can be written in
the following form:
\begin{equation}
\left. v_{\varphi }^{2}\right| _{z=0}=\frac{1}{rF_{r}}\zeta \left( r\right)
,\ \zeta \left( r\right) \equiv \tilde{\alpha}\left[ \left( 1+\frac{%
T_{\infty }r^{1+\tilde{\alpha}}}{\hat{T}-T_{\infty }}\right) \ln \left( 1+%
\frac{\hat{T}-T_{\infty }}{T_{\infty }r^{1+\tilde{\alpha}}}\right) -1\right]
.  \label{IF160}
\end{equation}
Considering $\tilde{\alpha}\geq 0$, it is easy to check that $\zeta \left(
r\right) $ is decreasing function of $r$. Hence, we obtain, in view of (\ref
{IF160}), that
\begin{equation}
0\leq \tilde{\alpha}\leq \tilde{\alpha}_{m}\equiv \left( \frac{\hat{T}}{\hat{%
T}-T_{\infty }}\ln \frac{\hat{T}}{T_{\infty }}-1\right) ^{-1}.  \label{IF170}
\end{equation}
Obviously, if $\tilde{\alpha}=\tilde{\alpha}_{m}$, then the midplane
circular velocity coincides with the Keplerian velocity at $r=1$.
Consequently, $\left. v_{\varphi }^{2}\right| _{z=0}$ will be close to $%
v_{\varphi k}^{2}$ in a small vicinity of the point $r=1$. If, however, we
assume that $\tilde{\alpha}=0$, then $v_{\varphi }\equiv 0$ in $\frak{R}_{a}$%
. Let us also show that there exists a subregion of $\frak{R}_{a}$, where
the midplane circular velocity differs significantly from the Keplerian
velocity even if $\tilde{\alpha}=\tilde{\alpha}_{m}$. We will use the
following asymptotic expansion in the limit $r\rightarrow \infty $:
\begin{equation}
\ln \left( 1+\frac{\hat{T}-T_{\infty }}{T_{\infty }r^{1+\tilde{\alpha}_{m}}}%
\right) =\frac{\hat{T}-T_{\infty }}{T_{\infty }r^{1+\tilde{\alpha}_{m}}}-%
\frac{1}{2}\left( \frac{\hat{T}-T_{\infty }}{T_{\infty }r^{1+\tilde{\alpha}%
_{m}}}\right) ^{2}+\ldots \ .  \label{IF180}
\end{equation}
We obtain, by virtue of (\ref{IF180}), the asymptotic expansion for $\left.
v_{\varphi }^{2}\right| _{z=0}$ in the limit $r\rightarrow \infty $:
\begin{equation}
\left. v_{\varphi }^{2}\right| _{z=0}=\frac{C_{\infty }^{2}}{F_{r}r^{2+%
\tilde{\alpha}_{m}}}+o\left( \frac{1}{r^{2+\tilde{\alpha}_{m}}}\right)
,\quad C_{\infty }^{2}=\tilde{\alpha}_{m}\frac{\hat{T}-T_{\infty }}{%
2T_{\infty }}.  \label{IF190}
\end{equation}
Thus, as we can see from (\ref{IF190}), the midplane circular velocity
differs significantly from the Keplerian one for sufficiently large $r$.

Thus, the simple, well-known models considered above clearly show that the
midplane circular velocity can be sub-Keplerian or even highly
non-Keplerian. Hence, the ``justification'' of the widely known assertion
that an inviscid (or laminar viscous) accretion disk must be sub-Keplerian
(see, e.g., \cite{Boss 2005}, \cite{Duric 2004}, \cite{Frank et al. 2002},
\cite{Fridman et al 2006}, \cite{Nature Publishing Group 2001}, \cite
{Shapiro and Teukolsky 2004}, \cite{Shore 2007}, \cite{Vietri Mario 2008})
is questionable. In this connection, mention should be made of another
approach, in which one simply assumes that radial pressure forces, i.e. the
first term on the RHS of (\ref{InvF160}), are negligible (see, e.g.,
\cite[p. 165]{Clarke and Carswell 2007}, \cite[p. 153]{Garcia 2011},
\cite[p. 132]{Hartmann 2009}). Hence, in view of (\ref{InvF160}), the disk
is almost Keplerian. The arguments for such a decision are based, to a large
extent, on the observations of accretion disks and the processes in them,
despite the fact that the motion of an inviscid liquid can differ
significantly from the turbulent flows that take place in real disks. The
surprising thing is that accretion disks are ``proven'' to be Keplerian in
zeroth approximation\footnote{%
In zeroth approximation only the first terms of asymptotic expansions are
taken into consideration.}. The ``proof'' was done (see, e.g., \cite
{Kluzniak and Kita 2000}, \cite{Liverts et al. 2012}, \cite{Ogilvie 1997},
\cite{Rebusco et al 2009}, \cite{Regev 1983}, \cite{Regev et al. 2016}, \cite
{Shtemler 2011}, \cite{Shtemler 2009}, \cite{Umurhan and Shaviv 2005}) by
asymptotic methods\footnote{%
One of the most effective approach for solving the problem (\ref{Int130})-(%
\ref{Int160}), i.e. equations of mathematical physics containing small
parameters, is to apply asymptotic methods (e.g., \cite{Il'in 1992}, \cite
{Mishchenko and Rozov 1980}, \cite{Moiseev 1981}, \cite{Vasil'eva et al.
1995}, \cite{Wasow 1987}).}, generally by the regular perturbation
technique, applied to singularly perturbed PDEs. A thorough explanation and
exemplification of the fundamental difference between regular and singular
perturbations can be found, e.g., in \cite{Moiseev 1981}, \cite{Vasil'eva et
al. 1995}. It turns out that the asymptotic solution to the singularly
perturbed PDEs, i.e., in general, the equations with small parameters
multiplying derivatives, consists of two parts, namely of two power series
in small parameters. One power series is the so called regular part (e.g.,
\cite{Samoylenko 2004}, \cite{Vasil'eva et al. 1995}), where the
coefficients of the series are functions of non-dependent variables. Another
part, where the coefficients of the power series are functions of stretched
variables, is to be considered as a singular part \cite{Samoylenko 2004} of
the asymptotic solution. This part is often called as the boundary layer
part (or series) of the asymptotic solution \cite{Vasil'eva et al. 1995}. N.
N. Moiseev \cite{Moiseev 1981} pointed out that it is not quite felicitous
term for the singular part, since the purpose of singular series (together
with the regular one) is not only to satisfy the imposed boundary conditions
(cf., e.g., \cite{Regev et al. 2016}). In many works (see, e.g., \cite
{Kluzniak and Kita 2000}, \cite{Liverts et al. 2012}, \cite{Rebusco et al
2009}, \cite{Regev 2008}, \cite{Shtemler 2011}, \cite{Shtemler 2009}, \cite
{Umurhan and Shaviv 2005}) the authors use only the regular part of
asymptotics, in spite of the fact that the PDEs being investigated are
singularly perturbed. Such an approach can lead to erroneous conclusions. To
demonstrate it, let us consider the system (\ref{InvF160})-(\ref{InvF170})
in the region $\frak{R}_{a}$ = $\left\{ \left( r,z\right) :\right. $ $r\in
\left[ 1,\infty \right) $, $\left. z\in \left[ 0,\infty \right) \right\} $.
We assume that $0$ $<$ $E_{u}$ $\ll $ $1$, the gas is ideal, (\ref{Int175}),
and the temperature $T=T\left( r,z\right) $ is a pre-assigned function of
the coordinates in the region $\frak{R}_{a}$.

We assume, at first, that the asymptotic solution to the system consists
only of the regular series, as it has been done in many astrophysical
publications dealing with the asymptotics. Let
\begin{equation*}
\rho =\rho _{0}+E_{u}\rho _{1}+E_{u}^{2}\rho _{2}+\ldots ,\quad
P=P_{0}+E_{u}P_{1}+E_{u}^{2}P_{2}+\ldots ,
\end{equation*}
\begin{equation}
v_{\varphi }=v_{\varphi 0}+E_{u}v_{\varphi 1}+E_{u}^{2}v_{\varphi 2}+\ldots ,
\label{IF200}
\end{equation}
where $\rho _{i}=\rho _{i}\left( r,z\right) ,$ $P_{i}=P_{i}\left( r,z\right)
,$ $v_{\varphi i}=v_{\varphi i}\left( r,z\right) ,$ $i=0,1,2,\ldots $ .
Then, substituting (\ref{IF200}) into (\ref{InvF160})-(\ref{InvF170}) and
equating coefficients of the same powers of $E_{u}$ in both sides of (\ref
{InvF160}) and (\ref{InvF170}), we obtain the problems for the terms $\rho
_{i}$, $P_{i}$, $v_{\varphi i}$, $i=0,1,2,\ldots $ . Specifically, we obtain
the following equations for zeroth approximation:
\begin{equation}
\frac{\rho _{0}v_{\varphi 0}^{2}}{r}=\frac{\rho _{0}}{F_{r}}\frac{\partial
\Phi }{\partial r},\quad \frac{\rho _{0}}{F_{r}}\frac{\partial \Phi }{%
\partial z}=0.  \label{IF210}
\end{equation}
By virtue of (\ref{Int165}) and (\ref{Int175}) we find from (\ref{IF210}):
\begin{equation*}
\rho _{0}=\left\{
\begin{array}{cc}
0, & z>0 \\
\rho _{0}^{\ast }, & z=0
\end{array}
\right. \ \Rightarrow \ P_{0}=\left\{
\begin{array}{cc}
0, & z>0 \\
P_{0}^{\ast }, & z=0
\end{array}
\right. ,
\end{equation*}
\begin{equation}
\left. v_{\varphi 0}\right| _{z=0}=\left\{
\begin{array}{cc}
v_{\varphi 0}^{\ast }, & \rho _{0}^{\ast }=0 \\
v_{\varphi k}\equiv 1\diagup \sqrt{rF_{r}}, & \rho _{0}^{\ast }\neq 0
\end{array}
\right. ,  \label{IF220}
\end{equation}
where $\rho _{0}^{\ast }$, $P_{0}^{\ast }$, and $v_{\varphi 0}^{\ast }$ are
arbitrary functions of $r$. Thus, in zeroth approximation, the disk is
Keplerian and its thickness is equal to zero. The equations for the first
approximation are the following:
\begin{equation}
\frac{2v_{\varphi 0}v_{\varphi 1}\rho _{0}+v_{\varphi 0}^{2}\rho _{1}}{r}=%
\frac{\partial P_{0}}{\partial r}+\frac{\rho _{1}}{F_{r}}\frac{\partial \Phi
}{\partial r},\quad \frac{\partial P_{0}}{\partial z}+\frac{\rho _{1}}{F_{r}}%
\frac{\partial \Phi }{\partial z}=0.  \label{IF230}
\end{equation}
Let $\iota $ be hyper-small but non-zero, i.e. $\iota $ $\simeq $ $0$. Then,
in view of (\ref{Int210}) and (\ref{IF220}), the second equation in (\ref
{IF230}) can be written as the following:
\begin{equation}
\frac{\partial P_{0}}{\partial z}+\frac{\rho _{1}}{F_{r}}\frac{\partial \Phi
}{\partial z}\equiv \left\{
\begin{array}{cc}
\frac{\rho _{1}}{F_{r}}\frac{\partial \Phi }{\partial z}=0, & z>0 \\
\frac{-P_{0}^{\ast }}{\iota }=0, & z=0
\end{array}
\right. ,  \label{IF240}
\end{equation}
whence $P_{0}^{\ast }=0$, and, in view of (\ref{IF220}), $P_{0}\equiv 0$.
Therefore we have, in zeroth approximation, that the radial pressure forces
are equal to zero. It proves the assumption about negligibility of the
radial pressure forces (see \cite[p. 165]{Clarke and Carswell 2007},
\cite[p. 153]{Garcia 2011}, \cite[p. 132]{Hartmann 2009}). Nontheless, we
continue our calculations. We find for high-order approximations:
\begin{equation}
\rho _{i}=\left\{
\begin{array}{cc}
0, & z>0 \\
\rho _{i}^{\ast }, & z=0
\end{array}
\right. ,\quad P_{i}\equiv 0,\quad \left. v_{\varphi i}\right| _{z=0}\equiv
0,\quad i=1,2,3,\ldots \,,  \label{IF250}
\end{equation}
where $\rho _{i}^{\ast }$ ($=\rho _{i}^{\ast }\left( r\right) $) is an
arbitrary function of $r$. Consequently, in view of (\ref{IF200}), we get
the following solution to the system (\ref{InvF160})-(\ref{InvF170}):
\begin{equation}
\rho =\left\{
\begin{array}{cc}
0, & z>0 \\
\rho ^{\ast }, & z=0
\end{array}
\right. ,\quad P\equiv 0,\quad \left. v_{\varphi }\right| _{z=0}=\left\{
\begin{array}{cc}
v_{\varphi }^{\ast }, & \rho ^{\ast }=0 \\
v_{\varphi k}\equiv 1\diagup \sqrt{rF_{r}}, & \rho ^{\ast }\neq 0
\end{array}
\right. ,  \label{IF260}
\end{equation}
where$\ \rho ^{\ast }=\rho _{0}^{\ast }+E_{u}\rho _{1}^{\ast }+E_{u}^{2}\rho
_{2}^{\ast }+\ldots $, $v_{\varphi }^{\ast }$ is an arbitrary function of $r$%
. Then, in view of (\ref{Int175}) and (\ref{IF260}), we obtain that $\rho
^{\ast }=0$ if $\left. T\right| _{z=0}\equiv T^{\circ }\left( r\right) \neq
0 $, otherwise, if $T^{\circ }\left( r\right) =0$ then $\rho ^{\ast }$ is an
arbitrary function of $r$. Thus, in the case when $T\equiv 0$ we obtain that
the disk is Keplerian, its thickness is equal to zero, the gas density is an
arbitrary function of $r$, and the disk consists of non-interacting
particles (since $P\equiv 0$). This result makes some sense if we are
dealing with say a pressureless dust . In the case when $T>0$, the
assumption that the solution to the system (\ref{Int175}), (\ref{InvF160})-(%
\ref{InvF170}) consists only of the regular series leads to the conclusion
that $\rho \equiv 0$ in the whole region $\frak{R}_{a}$ = $\left\{ \left(
r,z\right) :\right. $ $r\in \left[ 1,\infty \right) $, $\left. z\in \left[
0,\infty \right) \right\} $ and, hence, there is no any flow in $\frak{R}%
_{a} $ for lack of gas.

Now, let us assume that the asymptotic solution to the system (\ref{Int175}%
), (\ref{InvF160})-(\ref{InvF170}) consists of two parts \cite{Vasil'eva et
al. 1995}, namely regular and singular power series in a small parameter. We
will consider the case when $T=T\left( r,z\right) >0$ in $\frak{R}_{a}$. Let
us take advantage of the fact that the radial coordinate $r$ enters the
equation (\ref{InvF170}) as a parameter only. Then, in view of (\ref{Int175}%
), the equation (\ref{InvF170}) can be seen as singularly perturbed ODE and,
hence, it can be solved separately from Eq. (\ref{InvF160}). For the sake of
convenience, we will use, in general, the notation introduced in \cite
{Vasil'eva et al. 1995}. The stretched variable, $s$, is defined as follows
\begin{equation}
s=\frac{z}{\epsilon },\quad \epsilon \equiv \sqrt{E_{u}}.  \label{IF270}
\end{equation}
We will seek the asymptotic expansion for the solution of Eq. (\ref{InvF170}%
) in the following form.
\begin{equation}
P=\Lambda P\left( z,E_{u}\right) +\Pi P\left( s,\epsilon \right) ,
\label{IF280}
\end{equation}
where $\Lambda P\left( z,E_{u}\right) $ and $\Pi P\left( s,\epsilon \right) $%
\ are, respectively, the regular and singular parts of the expansion,
namely,
\begin{equation*}
\Lambda P=\Lambda _{0}P\left( z\right) +E_{u}\Lambda _{1}P\left( z\right)
+E_{u}^{2}\Lambda _{2}P\left( z\right) +\ldots ,
\end{equation*}
\begin{equation}
\Pi P=\Pi _{0}P\left( s\right) +\epsilon \Pi _{1}P\left( s\right) +\epsilon
^{2}\Pi _{2}P\left( s\right) +\ldots \,.  \label{IF290}
\end{equation}
By virtue of (\ref{Int165}) and (\ref{Int175}), we rewrite (\ref{InvF170})
to read:
\begin{equation}
E_{u}\frac{\partial P}{\partial z}=\digamma \left( P,T\left( r,z\right)
,r,z\right) \equiv -\frac{zP}{F_{r}T\left( r,z\right) \sqrt{\left(
r^{2}+z^{2}\right) ^{3}}},\ 0<E_{u}\ll 1.  \label{IF300}
\end{equation}
One can readily see, by virtue of Vasil'eva theorem \cite[p. 26]{Vasil'eva
et al. 1995}, that the series (\ref{IF290}) will be asymptotic in the
interval $z>0$ if the function $T^{-1}\left( r,z\right) $ will be infinitely
differentiable with respect to $z$ in the interval $z\geq 0$. Substituting
the series (\ref{IF290}) into Eq. (\ref{IF300}), we obtain:
\begin{equation}
E_{u}\frac{\partial \Lambda P}{\partial z}+\epsilon \frac{\partial \Pi P}{%
\partial s}=\digamma \left( \Lambda P+\Pi P,T\left( r,z\right) ,r,z\right) .
\label{IF310}
\end{equation}
Following \cite{Vasil'eva et al. 1995}, we rewrite the function $\digamma $
of Eq. (\ref{IF310}) in the form similar to (\ref{IF280}). Taking into
account the linear dependence of $\digamma $ on $P$, we obtain:
\begin{equation*}
\digamma \left( \Lambda P+\Pi P,T\left( r,z\right) ,r,z\right) =\digamma
\left( \Lambda P\left( z,E_{u}\right) ,T\left( r,z\right) ,r,z\right) +
\end{equation*}
\begin{equation}
\digamma \left( \Pi P\left( s,\epsilon \right) ,T\left( r,s\epsilon \right)
,r,s\epsilon \right) \equiv \Lambda \digamma +\Pi \digamma ,  \label{IF320}
\end{equation}
where $\Lambda \digamma $\ denotes the expansion of $\digamma \left( \Lambda
P,T\left( r,z\right) ,r,z\right) $ in terms of powers of $E_{u}$ with the
coefficients depending on $z$, and $\Pi \digamma $ denotes the expansion of
the rest part of $\digamma \left( \Lambda P+\Pi P,T\left( r,z\right)
,r,z\right) $ in powers of $\epsilon $ with the coefficients depending on $s$%
. Then, by virtue of (\ref{IF320}) and (\ref{IF290}), we obtain from (\ref
{IF310}) that
\begin{equation}
\frac{z\Lambda _{0}P}{F_{r}T\sqrt{\left( r^{2}+z^{2}\right) ^{3}}}=0,\,\frac{%
\partial P_{i-1}}{\partial z}=-\frac{z\Lambda _{i}P}{F_{r}T\sqrt{\left(
r^{2}+z^{2}\right) ^{3}}},\,i=1,2,\ldots \,.  \label{IF330}
\end{equation}
Since it is assumed that $T=T\left( r,z\right) >0$ in $\frak{R}_{a}$, we
obtain from (\ref{IF330}) that $\Lambda _{i}P\left( z\right) \equiv 0$ for $%
i=0,1,2,\ldots $ . Hence, $\Lambda P\equiv 0$ in the region $\frak{R}_{a}$.
From the above reasoning it is clear that the asymptotic solution to Eq. (%
\ref{InvF170}) consists only of the singular part. In view of (\ref{IF290}),
the equation (\ref{IF310}) can now be written as follows:
\begin{equation}
\frac{\partial \left[ \Pi _{0}P\left( s\right) +\epsilon \Pi _{1}P\left(
s\right) +\ldots \right] }{\partial s}=-\frac{s\left[ \Pi _{0}P\left(
s\right) +\epsilon \Pi _{1}P\left( s\right) +\ldots \right] }{F_{r}T\left(
r,\epsilon s\right) \sqrt{\left( r^{2}+\epsilon ^{2}s^{2}\right) ^{3}}}.
\label{IF340}
\end{equation}
Notice that $\Pi _{i}P\left( s\right) $ ($i=0,1,2,\ldots $) should be
boundary functions, namely, they should approach zero as $s\rightarrow
\infty $ \cite{Vasil'eva et al. 1995}. For the leading term of the singular
part we have:
\begin{equation}
\frac{\partial \Pi _{0}P\left( s\right) }{\partial s}=-\frac{s\Pi
_{0}P\left( s\right) }{F_{r}r^{3}T^{\circ }\left( r\right) },\quad T^{\circ
}\left( r\right) \equiv \left. T\left( r,z\right) \right| _{z=0}.
\label{IF350}
\end{equation}
Let $P^{\ast }\equiv \Pi _{0}P\left( 0\right) $. Then, in view of (\ref
{IF350}), we have that:
\begin{equation}
\Pi _{0}P\left( s\right) =P^{\ast }\exp \left\{ -\frac{s^{2}}{%
2F_{r}r^{3}T^{\circ }\left( r\right) }\right\} .  \label{IF360}
\end{equation}
Let us restrict our consideration to the zeroth approximation. We assume
that $P^{\ast }=P^{\circ }\equiv \left. P\left( r,z\right) \right| _{z=0}$.
Then, by virtue of (\ref{IF270}), (\ref{IF280}), and (\ref{IF360}), we
obtain:
\begin{equation}
P_{0}\equiv \Lambda _{0}P\left( z\right) +\Pi _{0}P\left( z\diagup \sqrt{%
E_{u}}\right) =P^{\circ }\left( r\right) \exp \left\{ -\frac{z^{2}}{%
2E_{u}F_{r}r^{3}T^{\circ }\left( r\right) }\right\} ,  \label{IF370}
\end{equation}
where $P^{\circ }\left( r\right) $ is an arbitrary function of $r$. Recall
that we consider the gas flow in the region $\frak{R}_{a}$=$\left\{ \left(
r,z\right) :\text{ }r\in \left[ 1,\infty \right) \text{, }z\in \left[
0,\infty \right) \right\} $ due to the axisymmetry of the problem, in spite
of the fact that the system (\ref{InvF160})-(\ref{InvF170}) is valid for all
values of $z\in \left( -\infty ,\infty \right) $. We have, because of the
axisymmetry, the additional condition for the pressure at $z=0$, namely $%
\left. \partial P\diagup \partial z\right| _{z=0}=0$, which is obviously
valid for (\ref{IF370}). Notice also that we cannot evaluate $P^{\circ
}\left( r\right) $ by virtue of the boundary condition (\ref{InvF60}), since
$\left. P_{0}\left( r,z\right) \right| _{z\rightarrow \infty }\rightarrow 0$
in (\ref{IF370}). Let us use the exact solution (\ref{IF100}) to Eq. (\ref
{InvF170}) for the evaluation of $P^{\circ }\left( r\right) $ in (\ref{IF370}%
). In such a case we obtain from (\ref{IF370}) that
\begin{equation}
P_{0}=P_{\infty }\exp \left\{ \frac{1}{E_{u}F_{r}}\left[ \int\limits_{0}^{%
\infty }\frac{\xi }{T\left( r,\xi \right) \left( r^{2}+\xi ^{2}\right) ^{1.5}%
}d\xi -\frac{z^{2}}{2r^{3}T^{\circ }\left( r\right) }\right] \right\} .
\label{IF380}
\end{equation}
Using the exact solution (\ref{IF100}) to Eq. (\ref{InvF170}), we find the
relative error $\delta _{P}\equiv \left| \left( P-P_{0}\right) \diagup
P\right| $:
\begin{equation}
\delta _{P}=\left| 1-\exp \left\{ \frac{1}{E_{u}F_{r}}\left[
\int\limits_{0}^{z}\frac{\xi }{T\left( r,\xi \right) \left( r^{2}+\xi
^{2}\right) ^{1.5}}d\xi -\frac{z^{2}}{2r^{3}T^{\circ }\left( r\right) }%
\right] \right\} \right| .  \label{IF385}
\end{equation}
Obviously, the value of $\delta _{P}$ will be comparatively small in a small
vicinity of $z=0$ (in the boundary layer \cite[p. 13]{Vasil'eva et al. 1995}%
), since $E_{u}F_{r}\ll 1$. In this connection, it should be noted that the
singular part of the asymptotic expansion is usually necessary for correct
solution of singularly perturbed problems like (\ref{InvF160})-(\ref{InvF170}%
).

\subsubsection{Instability of inviscid disk\label%
{Instability of inviscid disk}}

We consider the axisymmetric steady-state flows\footnote{%
Recall that the flows considered in this section are assumed to be laminar
and inviscid.} in the region $\frak{R}_{a}$ = $\left\{ \left( r,z\right)
:\right. $ $r\in \left[ 1,\infty \right) $, $\left. z\in \left[ 0,\infty
\right) \right\} $. If the process in question is polytropic, i.e. (\ref
{Int176}) is valid, then this process is described by the balance equations,
(\ref{InvF10})-(\ref{InvF40}), as well as by the equation of state:
\begin{equation}
P=P\left( \rho ,T\right) ,\quad \left. P\right| _{T=0}=0.  \label{IID1010}
\end{equation}
For instance, if the gas is perfect, then Eq. (\ref{IID1010}) is reduced to
Eq. (\ref{Int175}). If the process is non-polytropic, then the steady-state
axisymmetric version of the energy balance equation, (\ref{Int30}), should
be used instead of Eq. (\ref{Int176}). We will mainly consider the ODE
system, (\ref{InvF80}) - (\ref{InvF100}), in the midplane variables $%
\varsigma ^{\circ }\equiv \left. \varsigma \right| _{z=0}$, where $\varsigma
=\varsigma \left( r,z\right) $ denotes a dependent variable. It is assumed
that $q^{\circ }\left( r\right) \equiv r\rho ^{\circ }v_{r}^{\circ }$ is
continuous (e.g. \cite[p. 106]{Ponstein 2001}) at all $r\in \left[ 1,\infty
\right) $.

First of all we prove that any non-vortex flow\footnote{%
Recall that an axisymmetric flow with $v_{\varphi }\varpropto r^{-1}$, (\ref
{InvF110}), is to be called as vortex. In particular, vortex-sinks and
vortex-sources are lumped together as vortices.
\par
{}} is unstable. We consider the steady-state accretion disk as input-output
\cite{Harris 1983} (or cause-effect \cite{Mesarovic and Takahara 1975})
system. The boundary conditions, namely (\ref{InvF50}), (\ref{InvF60}), as
well as
\begin{equation}
\left. q\right| _{r=1}=Q,\quad Q=Q\left( z\right) ,\ \left. Q\left( z\right)
\right| _{z\rightarrow \infty }\rightarrow 0,\ q\equiv r\rho v_{r},
\label{IID1015}
\end{equation}
\begin{equation}
\left. v_{\varphi }\right| _{r=1}=V,\quad V=V\left( z\right) ,\ \left.
V\left( z\right) \right| _{z\rightarrow \infty }\rightarrow 0,
\label{IID1020}
\end{equation}
are considered as inputs and the vector-functions $\left\{ \rho \left(
r,z\right) ,\right. $ $P\left( r,z\right) ,$ $T\left( r,z\right) ,$ $%
v_{r}\left( r,z\right) ,$ $v_{\varphi }\left( r,z\right) ,$ $\left.
v_{z}\left( r,z\right) \right\} $ are considered as outputs. Recall that we
consider the gas flow in the region $\frak{R}_{a}$ = $\left\{ \left(
r,z\right) :\right. $ $r\in \left[ 1,\infty \right) $, $\left. z\in \left[
0,\infty \right) \right\} $ due to the axisymmetry and, hence, we have:

\begin{equation}
\left. v_{z}\right| _{z=0}=0.  \label{IID1030}
\end{equation}
The mapping $F$: input $\rightarrow $ output is defined by the steady-state
balance equations, (\ref{InvF10})-(\ref{InvF40}), the steady-state
axisymmetric version of the energy balance equation, (\ref{Int30}), as well
as by the equation of state, (\ref{IID1010}). Following, e.g., \cite{Harris
1983}, \cite{Lurie and Enright 2012}, and \cite{Mesarovic and Takahara 1975}
we can say that the input-output system is unstable if an infinitesimal
increment in the input triggers a finite increment in the output. Such an
approach was already used, \cite{Borisov 2013}, to investigate instability
of magnetic as well as non-magnetic accretion disks. In our case this
approach gives the result in a few simple steps. For instance, it is assumed
(see \cite{Borisov 2013}) that $v_{r}=0$ to have a possibility for $%
v_{\varphi }$ be, e.g., sub-Keplerian. Then, as the input, we take an
infinitesimal increment in $v_{r}$. In such a case we obtain that $%
v_{\varphi }$ will be a vortex, i.e., we obtain a finite increment in the
output. In this subsection the instability of gas-dynamic (non-magnetic)
accretion disks will be considered in more details. It is already shown
above (see Subsection \ref{Circular velocity}) that the circular velocity
may differ from the vortex velocity only if the radial velocity is equal to
zero. Accordingly, we assume that $q\equiv r\rho v_{r}=0$ ($st\left( \rho
\right) >0$)\ at all $\left( r,z\right) \in \frak{R}_{a}$ and, hence,
\begin{equation}
Q=0,\quad \forall z\in \left[ 0,\infty \right) .  \label{IID1035}
\end{equation}
Let $v_{\varphi }^{\circ }=v_{\varphi }^{\circ }\left( r\right) $ be a
certain midplane circular velocity. For instance, we may assume that the
circular velocity will be ``Keplerian with a great precision'', as it is
argued in, e.g, \cite{Clarke and Carswell 2007}, \cite{Frank et al. 2002},
\cite{Fridman et al 2006}, \cite{Garcia 2011}, \cite{Hartmann 2009}, \cite
{Shore 2007}, \cite{Vietri Mario 2008}. An important point is that $%
v_{\varphi }^{\circ }=v_{\varphi }^{\circ }\left( r\right) $ is assumed to
be significantly different from the vortex velocity $v_{\varphi v}^{\circ
}=C_{\varphi }^{\circ }\diagup r$, i.e. for any interval $(a,b)\subset \frak{%
R}_{a}^{\circ }$ $\equiv $ $\left\{ \left( r,z\right) :\right. $ $r\in \left[
1,\infty \right) $, $\left. z=0\right\} $ with $a\ncong b$ we have
\begin{equation}
\left\| v_{\varphi }^{\circ }-v_{\varphi v}^{\circ }\right\| _{\infty
,(a,b)}\equiv \left\| v_{\varphi }^{\circ }-\frac{C_{\varphi }^{\circ }}{r}%
\right\| _{\infty ,(a,b)}\ncong 0,\quad \forall \,C_{\varphi }^{\circ
}=const.  \label{IID1040}
\end{equation}
Here $\left\| f\right\| _{\infty }=\left\| f\right\| _{\infty ,S}\equiv \sup
\left\{ \left| f\left( r\right) \right| :r\in S\right\} $ denotes the
Chebyshev norm of a bounded function $f\left( r\right) $ defined on a set $S$%
. Let us change the value of $Q$, (\ref{IID1035}), in the boundary condition
(\ref{IID1015}), namely, let $Q\rightarrow \widetilde{Q}$:
\begin{equation}
\widetilde{Q}=\iota \widetilde{a}_{q},\quad \iota =const\simeq 0,\
\widetilde{a}_{q}=\widetilde{a}_{q}\left( z\right) ,\ \widetilde{a}%
_{q}^{\circ }\equiv \left. \widetilde{a}_{q}\right| _{z=0}\neq 0,
\label{IID1050}
\end{equation}
where the bounded function $\widetilde{a}_{q}\left( z\right) $ is such that $%
\left. \widetilde{a}_{q}\right| _{z\rightarrow \infty }\rightarrow 0$. All
other boundary conditions are unchanged. The change of the boundary
condition leads to the modification of the output, i.e. $\varsigma ^{\circ
}\rightarrow \widetilde{\varsigma }^{\circ }$, where $\varsigma $ denotes a
dependent variable. For instance, we obtain from (\ref{InvF80}) and (\ref
{IID1050}) that
\begin{equation}
q^{\circ }\equiv 0\rightarrow \tilde{q}^{\circ }=\iota \widetilde{a}%
_{q}^{\circ }-\int\limits_{1}^{r}\xi \tilde{\rho}^{\circ }\tilde{v}%
_{z}^{\prime \circ }d\xi ,\quad \widetilde{\rho }^{\circ }\equiv \left.
\widetilde{\rho }\right| _{z=0},\ \tilde{v}_{z}^{\prime \circ }\equiv \left.
\frac{\partial \tilde{v}_{z}}{\partial z}\right| _{z=0}.  \label{IID1060}
\end{equation}
Since $\tilde{q}^{\circ }=\tilde{q}^{\circ }\left( r\right) $ is continuous
at all $r\in \left[ 1,\infty \right) $ and $\left. \tilde{q}^{\circ }\right|
_{r=1}=\iota \widetilde{a}_{q}^{\circ }\neq 0$, (\ref{IID1050}), then there
exists a real number $r_{\ast }>1$ such that$\ \tilde{q}^{\circ }\neq 0$ at
all $r\in \left( 1,r_{\ast }\right) $. We find from (\ref{InvF100}) and (\ref
{IID1020}) that
\begin{equation}
\widetilde{v}_{\varphi }^{\circ }=\frac{C_{\varphi }^{\circ }}{r},\quad
\forall r\in \left[ 1,r_{\ast }\right) ,\ C_{\varphi }^{\circ }=const=\left.
V\right| _{z=0}.  \label{IID1070}
\end{equation}
Thus, we obtain that the infinitesimal increment in the input, i.e. $Q$ $%
\rightarrow $ $\widetilde{Q}$, see (\ref{IID1035}) and (\ref{IID1050}),
triggers the finite increment in the output (i.e. $v_{\varphi }^{\circ }$ $%
\rightarrow $ $\widetilde{v}_{\varphi }^{\circ }$), since $\left\|
v_{\varphi }^{\circ }-\widetilde{v}_{\varphi }^{\circ }\right\| _{\infty
,(1,r_{\ast })}\ncong 0$ in view of (\ref{IID1040}), which is to say that
the flow is unstable.

For the sake of illustration, let us consider the axisymmetric steady-state
flow with the circular velocity represented by the power-law model \cite[p.
374]{Vietri Mario 2008}. First, we assume that
\begin{equation}
q\equiv r\rho v_{r}=0,\quad \forall \left( r,z\right) \in \frak{R}_{a},\
st\left( \rho \right) >0,  \label{IID1075}
\end{equation}
and the midplane circular velocity, $v_{\varphi }^{\circ }$, is represented
as the following:

\begin{equation}
v_{\varphi }^{\circ }=\frac{C_{\varphi }^{\circ }}{r^{\varkappa }},\quad
\forall r\in \frak{R}_{a}^{\circ },\quad C_{\varphi }^{\circ }=const,\
0.5\leq \varkappa =const<1,  \label{IID1080}
\end{equation}
where $\frak{R}_{a}^{\circ }$ $\equiv $ $\left\{ \left( r,z\right) :\right. $
$r\in \left[ 1,\infty \right) $, $\left. z=0\right\} $. We emphasize that $%
st\left( \varkappa \right) <1$, i.e. the midplane circular velocity given by
(\ref{IID1080}) is assumed to be significantly different from the vortex
velocity $v_{\varphi v}^{\circ }=C_{\varphi }^{\circ }\diagup r$. The
boundary conditions are the same as above, namely: (\ref{InvF50}), (\ref
{InvF60}), (\ref{IID1015}), and (\ref{IID1020}). In view of (\ref{IID1075})
and (\ref{IID1080}), we have (\ref{IID1035}) and, respectively
\begin{equation}
\left( \left. v_{\varphi }^{\circ }\right| _{r=1}=\right) \ V^{\circ }\equiv
\left. V\right| _{z=0}=C_{\varphi }^{\circ }.  \label{IID1120}
\end{equation}
By virtue of (\ref{IID1075}) and (\ref{IID1030}) we obtain from \ref{InvF10}
that $v_{z}\equiv 0$ in the region $\frak{R}_{a}$. Recall that the condition
(\ref{IID1075}), i.e. $v_{r}\equiv 0$ in $\frak{R}_{a}$, was taken to
examine a non-vortex flow, which is possible only if the radial velocity is
equal to zero. If, however, we assume that $v_{z}\equiv 0$ in $\frak{R}_{a}$%
, then $v_{r}\equiv 0$ in $\frak{R}_{a}$, provided (\ref{IID1035}).

Let us change the value of $Q$, (\ref{IID1035}), in the boundary condition (%
\ref{IID1015}), namely, let $Q\rightarrow \tilde{Q}$:
\begin{equation}
\tilde{Q}=\left\{
\begin{array}{cc}
\iota \widetilde{a}_{q}\neq 0, & z=0 \\
0 & z>0
\end{array}
\right. ,\quad \iota \simeq 0.  \label{IID1130}
\end{equation}
Given $\tilde{v}_{z}\equiv 0$ in the region $\frak{R}_{a}$, we obtain from
\ref{InvF10}, in view of (\ref{IID1130}), that
\begin{equation}
\tilde{q}\left( r,z\right) \equiv r\widetilde{\rho }\tilde{v}_{r}=\left\{
\begin{array}{cc}
\iota \widetilde{a}_{q}\neq 0, & z=0 \\
0 & z>0
\end{array}
\right. ,\quad r\in \left[ 1,\infty \right) ,\ \iota \simeq 0.
\label{IID1140}
\end{equation}
We find from (\ref{InvF100}) and (\ref{IID1140}) that
\begin{equation}
\widetilde{v}_{\varphi }^{\circ }=\frac{\tilde{C}_{\varphi }^{\circ }}{r}%
,\quad \forall r\in \left[ 1,\infty \right) ,\ \tilde{C}_{\varphi }^{\circ
}=const.  \label{IID1150}
\end{equation}
We assume that $\tilde{C}_{\varphi }^{\circ }=C_{\varphi }^{\circ }$, see (%
\ref{IID1150}) and (\ref{IID1120}). In such a case $\left. \widetilde{v}%
_{\varphi }^{\circ }\right| _{r=1}=\left. v_{\varphi }^{\circ }\right|
_{r=1} $. Thus, we obtain that the infinitesimal increment in the input ($Q$
$\rightarrow $ $\tilde{Q}$), i.e. $st\left( \tilde{Q}_{\ast }-Q_{\ast
}\right) =0$, triggers the finite increment in the output ($v_{\varphi
}^{\circ }$ $\rightarrow $ $\widetilde{v}_{\varphi }^{\circ }$), since the
midplane circular velocity given by (\ref{IID1080}) is assumed to be
significantly different from the vortex velocity (\ref{IID1050}). Hence, the
flow in question is unstable with respect to infinitesimal perturbations in
the input (\ref{IID1015}).

Let us now examine the stability of the regular part of asymptotic solution,
(\ref{IF260}), to the singularly perturbed system (\ref{InvF10})-(\ref
{InvF40}), provided $v_{r}\equiv 0$ and, hence, $v_{z}\equiv 0$ in the
region $\frak{R}_{a}$. Recall that in the case when $T>0$, the assumption
that the solution consists only of the regular series, (\ref{IF200}), leads
to the conclusion that $\rho \equiv 0$ in the whole region $\frak{R}_{a}$
and, hence, there is no any flow in the whole region $\frak{R}_{a}$ for lack
of gas. Otherwise, i.e.\ $T\equiv 0$, we obtain that the disk is Keplerian (%
\ref{InvF114}), its thickness is equal to zero (i.e. $\rho =0$ for all $z>0$%
), the gas density at the midplane is an arbitrary function of $r$, and the
disk consists of non-interacting particles (since $P\equiv 0$). We will
assume that $st\left( \rho \right) >0$ for all $z=0$. It is easy to see
that this solution satisfies the degenerate system (see, e.g., \cite
{Mishchenko and Rozov 1980}, \cite{Vasil'eva et al. 1995}) corresponding to (%
\ref{InvF10})-(\ref{InvF40}), i.e. the system obtained from (\ref{InvF10})-(%
\ref{InvF40}) by putting $E_{u}=0$. The system (\ref{InvF10})-(\ref{InvF40})
at the midplane ($z=0$) becomes the following ODE system in the midplane
variables.
\begin{equation}
\frac{1}{r}\frac{\partial \left( r\rho ^{\circ }v_{r}^{\circ }\right) }{%
\partial r}=0,  \label{IID1160}
\end{equation}
\begin{equation}
\frac{\partial }{\partial r}\left[ r\rho ^{\circ }\left( v_{r}^{\circ
}\right) ^{2}\right] -\rho ^{\circ }\left( v_{\varphi }^{\circ }\right)
^{2}=-\frac{\rho ^{\circ }}{F_{r}r},  \label{IID1170}
\end{equation}
\begin{equation}
r\rho ^{\circ }v_{r}^{\circ }\frac{\partial rv_{\varphi }^{\circ }}{\partial
r}=0.  \label{IID1180}
\end{equation}
The boundary conditions are the following:
\begin{equation}
\left. q^{\circ }\right| _{r=1}=Q^{\circ },\quad q^{\circ }\equiv r\rho
^{\circ }v_{r}^{\circ },\quad \left. v_{r}^{\circ }\right| _{r\rightarrow
\infty }\rightarrow 0,  \label{IID1190}
\end{equation}
\begin{equation}
\left. v_{\varphi }^{\circ }\right| _{r=1}=V^{\circ },\quad \left.
v_{\varphi }^{\circ }\right| _{r\rightarrow \infty }\rightarrow 0,\quad
\left. \rho ^{\circ }\right| _{r\rightarrow \infty }\rightarrow \rho
_{\infty }^{\circ }=const>0.  \label{IID1200}
\end{equation}
Obviously, if $Q^{\circ }=0$ in (\ref{IID1190}), then, in view of (\ref
{IID1160}), $q^{\circ }\equiv r\rho ^{\circ }v_{r}^{\circ }=0$ at all $r\in %
\left[ 1,\infty \right) $ and, hence, we find, by virtue of (\ref{IID1170}),
that the disk is Keplerian (\ref{InvF114}). It is possible if $V^{\circ
}=1\diagup \sqrt{F_{r}}$ in (\ref{IID1200}). Let us change the values of $%
Q^{\circ }$ and $V^{\circ }$ in the boundary conditions (\ref{IID1190}) and,
respectively, (\ref{IID1200}), namely, let $Q^{\circ }\rightarrow \tilde{Q}%
^{\circ }$ and $V^{\circ }\rightarrow \tilde{V}^{\circ }$:
\begin{equation}
\tilde{Q}^{\circ }=\iota \widetilde{a}_{q},\quad \iota \simeq 0,\ \widetilde{%
a}_{q}=const\neq 0,\quad st\left( \tilde{V}^{\circ }\right) =st\left(
V^{\circ }\right) .  \label{IID1210}
\end{equation}
In view of (\ref{IID1210}), we obtain from (\ref{IID1160}) that $\tilde{q}%
^{\circ }\equiv r\tilde{\rho}^{\circ }\tilde{v}_{r}^{\circ }=\iota
\widetilde{a}_{q}$ at all $r\in \left[ 1,\infty \right) $. Since $V^{\circ
}\rightarrow \tilde{V}^{\circ }$, we find from (\ref{IID1180}), by virtue of
(\ref{IID1200}), that
\begin{equation}
\widetilde{v}_{\varphi }^{\circ }=\frac{\tilde{V}^{\circ }}{r},\quad \forall
r\in \left[ 1,\infty \right) ,\ \tilde{V}^{\circ }=const.  \label{IID1220}
\end{equation}
Notice that the vortex velocity (\ref{IID1220}) does not satisfy Eq. (\ref
{IID1170}), since $\tilde{q}^{\circ }\equiv r\tilde{\rho}^{\circ }\tilde{v}%
_{r}^{\circ }=\iota \widetilde{a}_{q}$. Obviously, $st\left( \tilde{q}%
^{\circ }\right) $ $=$ $st\left( \iota \widetilde{a}_{q}\right) $ $=$ $%
st\left( \iota \right) st\left( \widetilde{a}_{q}\right) $ $=$ $0$. For the
sake of convenience, we assume that $st\left( r\right) =r$, $st\left(
F_{r}\right) =F_{r}$, and $st\left( F_{r}^{-1}\right) \neq 0$. It is also
assumed above that $st\left( \tilde{\rho}^{\circ }\right) \neq 0$.
Consequently, $st\left( r\tilde{\rho}^{\circ }\tilde{v}_{r}^{\circ }\right) $
$=$ $rst\left( \tilde{\rho}^{\circ }\right) st\left( \tilde{v}_{r}^{\circ
}\right) $ $=$ $0$ and, hence,
\begin{equation}
st\left( \tilde{v}_{r}^{\circ }\right) =0.  \label{IID1230}
\end{equation}
By virtue of (\ref{IID1160}), (\ref{IID1170}), (\ref{IID1190}), and (\ref
{IID1220}), we find:
\begin{equation*}
\tilde{v}_{r}^{\circ }\frac{\partial \tilde{v}_{r}^{\circ }}{\partial r}=%
\frac{\left( \tilde{V}^{\circ }\right) ^{2}}{r^{3}}-\frac{1}{F_{r}r^{2}}%
\quad \Rightarrow
\end{equation*}
\begin{equation}
\frac{\left( \tilde{v}_{r}^{\circ }\right) ^{2}}{2}=\frac{1}{F_{r}r}-\frac{%
\left( \tilde{V}^{\circ }\right) ^{2}}{2r^{2}},\quad \forall r\in \left[
1,\infty \right) .  \label{IID1240}
\end{equation}
It is easily seen that the equality (\ref{IID1240}) cannot be valid for all $%
r\in \left[ 1,\infty \right) $ due to (\ref{IID1230}). Actually, we obtain
from (\ref{IID1240}), by virtue of (\ref{IID1230}), that
\begin{equation}
\frac{1}{F_{r}}-\frac{\left( V^{\circ }\right) ^{2}}{2r}=0,\quad \forall
r\in \left[ 1,\infty \right) ,  \label{IID1250}
\end{equation}
where $V^{\circ }=st\left( \tilde{V}^{\circ }\right) =1\diagup \sqrt{F_{r}}$%
. Now it is clear that the equality (\ref{IID1250}) and, hence, the equality
(\ref{IID1240}) cannot be valid for all $r\in \left[ 1,\infty \right) $.
Consequently, the vortex velocity (\ref{IID1220}) does not satisfy Eq. (\ref
{IID1170}). Thus, the infinitesimal increment in the input ($Q^{\circ
}\rightarrow \tilde{Q}^{\circ }$, $V^{\circ }\rightarrow \tilde{V}^{\circ }$%
) does not trigger a finite increment in the output and, hence, the motion
in question is not unstable. An important remark is in order at this point.
The assumption that the asymptotic solution to the singularly perturbed PDEs
consists only of the regular series (such as (\ref{IF200}), see, e.g., \cite
{Kluzniak and Kita 2000}, \cite{Liverts et al. 2012}, \cite{Rebusco et al
2009}, \cite{Regev 2008}, \cite{Shtemler 2011}, \cite{Shtemler 2009}, \cite
{Umurhan and Shaviv 2005}) can lead to a wrong conclusion about the
stability of the system.

So, we have proven that any axisymmetric steady-state flow (excluding the
vortex (\ref{InvF110})) is unstable. The proof that the vortex motion is
unstable can be found in \cite{Borisov 2013} (see also \cite[Sec. 3.4]
{Tassoul 2000}, \cite[Sec. 7.2.4]{Vietri Mario 2008}).

\subsection{Viscous flow\label{Viscous disk}}

So far the inviscid model of accretion disk was examined. It has been proven
that the vortex, (\ref{InvF110}), will be the only exact solution for the
circular velocity field of midplane flow if the radial velocity $%
v_{r}^{\circ }\neq 0$. If $v_{r}^{\circ }=0$, then there exists a wide range
of solutions (including Keplerian and sub-Keplerian rotations) for the
midplane circular velocity. It has also been proven that any steady-state
solution to this model that does not coincide with the vortex will be
unstable (see Sec. \ref{Instability of inviscid disk}). The vortex will be
linearly unstable \cite{Borisov 2013} in the region under consideration. It
is also vital to note that the majority of observed disks are in Keplerian
(or sub-Keplerian) rotation around their central accreting objects (see,
e.g., \cite{Harsono et al. 2013}, \cite{Hure et al. 2011}, \cite{Murillo et
al. 2013}, \cite{Simon et al. 2001}, \cite{Takakuwa et al. 2012}, \cite{Yen
et al. 2014} and references therein). Since we consider the gas flows, it is
easy to show that the laminar viscous disk can be approximated with a great
precision by the vortex motion, (\ref{InvF110}).

The steady-state version of the PDE system (\ref{Int140})-(\ref{Int160}) is
thus reduced to the following one.

The $r$-component:
\begin{equation*}
\frac{1}{r}\frac{\partial }{\partial r}r\left( \rho v_{r}^{2}\right) +\frac{%
\partial }{\partial z}\left( \rho v_{r}v_{z}\right) -\frac{\rho v_{\varphi
}^{2}}{r}=-E_{u}\frac{\partial P}{\partial r}-\frac{\rho }{F_{r}}\frac{%
\partial \Phi }{\partial r}+
\end{equation*}
\begin{equation*}
\frac{1}{R_{e}}\left\{ \frac{\partial }{\partial r}\left[ 2\mu _{v}\frac{%
\partial v_{r}}{\partial r}-\frac{2}{3}\mu _{v}\left( \frac{1}{r}\frac{%
\partial \left( rv_{r}\right) }{\partial r}+\frac{\partial v_{z}}{\partial z}%
\right) \right] +\right.
\end{equation*}
\begin{equation}
\left. \frac{\partial }{\partial z}\mu _{v}\left( \frac{\partial v_{r}}{%
\partial z}+\frac{\partial v_{z}}{\partial r}\right) +\frac{2\mu _{v}}{r}%
\left( \frac{\partial v_{r}}{\partial r}\mathbf{-}\frac{v_{r}}{r}\right)
\right\} ,  \label{VF06}
\end{equation}
The $\varphi $-component (after simple transformations):
\begin{equation}
\frac{1}{r^{3}}\frac{\partial }{\partial r}\left( r^{3}\rho v_{r}\omega
\right) +\frac{\partial }{\partial z}\left( \rho v_{z}\omega \right) =\frac{1%
}{r^{3}R_{e}}\frac{\partial }{\partial r}\left( \mu _{v}r^{3}\frac{\partial
\omega }{\partial r}\right) +\frac{1}{R_{e}}\frac{\partial }{\partial z}%
\left( \mu _{v}\frac{\partial \omega }{\partial z}\right) ,  \label{VF07}
\end{equation}
The $z$-component:
\begin{equation*}
\frac{1}{r}\frac{\partial }{\partial r}r\left( \rho v_{z}v_{r}\right) +\frac{%
\partial }{\partial z}\left( \rho v_{z}^{2}\right) =
\end{equation*}
\begin{equation*}
-E_{u}\frac{\partial P}{\partial z}-\frac{\rho }{F_{r}}\frac{\partial \Phi }{%
\partial z}+\frac{1}{R_{e}}\left\{ \frac{\partial }{\partial r}\left[ \mu
_{v}\left( \frac{\partial v_{r}}{\partial z}+\frac{\partial v_{z}}{\partial r%
}\right) \right] +\right.
\end{equation*}
\begin{equation}
\left. \frac{\partial }{\partial z}\left[ 2\mu _{v}\frac{\partial v_{z}}{%
\partial z}-\frac{2}{3}\mu _{v}\left( \frac{1}{r}\frac{\partial \left(
rv_{r}\right) }{\partial r}+\frac{\partial v_{z}}{\partial z}\right) \right]
+\frac{\mu _{v}}{r}\left( \frac{\partial v_{r}}{\partial z}+\frac{\partial
v_{z}}{\partial r}\right) \right\} ,  \label{VF08}
\end{equation}
where $\Phi =-1\diagup \sqrt{r^{2}+z^{2}}$, $\omega $ denotes the angular
velocity, i.e. $v_{\varphi }=\omega r$.

Much attention is given in Sec. \ref{Circular velocity} to the investigation
of inviscid flow, (\ref{InvF10})-(\ref{InvF40}), on condition that $%
v_{r}\equiv 0$, namely Eqs. (\ref{InvF160})-(\ref{InvF170}) in the region $%
\frak{R}_{a}$ = $\left\{ \left( r,z\right) :\text{ }r\in \left[ 1,\infty
\right) \text{, }z\in \left[ 0,\infty \right) \right\} $. In the case of
viscous flow (provided $v_{r}\equiv 0$) we also have (\ref{InvF160})-(\ref
{InvF170}), instead of (\ref{VF06}), (\ref{VF08}), and the following $%
\varphi $-component:
\begin{equation}
\frac{1}{r^{3}}\frac{\partial }{\partial r}\left( \mu _{v}r^{3}\frac{%
\partial \omega }{\partial r}\right) +\frac{\partial }{\partial z}\left( \mu
_{v}\frac{\partial \omega }{\partial z}\right) =0,\quad v_{\varphi }=\omega
r.  \label{VF09}
\end{equation}
It is readily seen\ that we obtain, in general, analogous results to the
inviscid problem. To take an example, for the isothermal flow ($T\equiv
T_{\infty }=const>0$) of ideal gas, the circular velocity $v_{\varphi
}\equiv 0$ in $\frak{R}_{a}$. As a further example, assume that the
asymptotic solution to the system consists only of the regular series, (\ref
{IF200}). If $T>0$, then we obtain (see Sec. \ref{Circular velocity}) that $%
\rho \equiv 0$ in the whole region $\frak{R}_{a}$. Hence, any solution, $%
\omega =\omega \left( r,z\right) $, to Eq. (\ref{VF09}) is admissible for
lack of gas. If, however, $T\equiv 0$, then we obtain that the disk is
Keplerian, its thickness is equal to zero, the gas density is an arbitrary
function of $r$, and the disk consists of non-interacting particles (see
Sec. \ref{Circular velocity}). Notice that $\left. \mu _{v}\right| _{T=0}=0$%
, in view of (\ref{Int54}) or (\ref{Int56}), and, hence, Eq. (\ref{VF09})
will also be satisfied.

\subsubsection{Average equations\label{Average equations}}

Let us now use the the well known approach for the investigation the
accretion disk dynamics (see, e.g., \cite[Sec. 12.1]{Clarke and Carswell
2007}). We will average Eqs. (\ref{VF07}) and (\ref{InvF10}) over the depth
of disk. In the following we will sometimes assume that $T_{H}\equiv \left.
T\left( r,z\right) \right| _{z=H}\approx 0$ and $\rho _{H}\equiv \left. \rho
\left( r,z\right) \right| _{z=H}\approx 0$, where $H=H\left( r\right) $
denotes the disk semi-thickness. Let $z=H\left( r\right) $ be the
streamline. For the sake of convenience, we will use the following scheme:
\begin{equation}
\rho =\left\{
\begin{array}{cc}
\rho \left( r,z\right) >0, & \left| z\right| <H \\
0, & \left| z\right| \geq H
\end{array}
,\right.  \label{PG30}
\end{equation}
and
\begin{equation}
T=0,\ P=0,\quad \left| z\right| \geq H.  \label{PG35}
\end{equation}
The following notation will be used
\begin{equation}
\overline{f}\equiv \frac{1}{2H}\int\limits_{-H}^{H}f\left( r,z\right) dz,
\label{VF40}
\end{equation}
where $f\left( r,z\right) $ denotes an integrable function, $H=H\left(
r\right) $ denotes the disk semi-thickness. With the notation (\ref{VF40}),
we obtain:
\begin{equation}
\overline{\frac{\partial f}{\partial r}}=\frac{1}{2H}\frac{\partial }{%
\partial r}\left( 2H\overline{f}\right) -\frac{f_{z=H}+f_{z=-H}}{2H}\frac{%
\partial H}{\partial r}.  \label{VF42}
\end{equation}
It is assumed, (\ref{Int205}), that $\left| \partial H\diagup \partial
r\right| \ll 1$. In such a case the last term in (\ref{VF42}) can be seen as
negligible provided $\left| \left. f\right| _{z=\pm H}\right| $ is small
enough. We will consider the case when $\left. f\right| _{z=\pm H}\approx 0$%
, i.e.
\begin{equation}
\overline{\frac{\partial f}{\partial r}}\approx \frac{1}{2H}\frac{\partial }{%
\partial r}\left( 2H\overline{f}\right) .  \label{VF45}
\end{equation}
We will use the well known (and natural for thin disks) assumption that the
variation of $v_{\varphi }$ (and, hence, $\omega $)\ with $z$ is negligible
(see, e.g., \cite{Frank et al. 2002}). If $f\left( z\right) $ does not
change sign on the interval $\left( -H,H\right) $, then this assumption
leads to the following equality.
\begin{equation}
\int\limits_{-H}^{H}f\omega dz=\left. \omega \right| _{z=\xi
}\int\limits_{-H}^{H}fdz\approx \overline{\omega }\int\limits_{-H}^{H}fdz,
\label{VF50}
\end{equation}
where $\xi \in \left[ -H,H\right] $. In a similar manner we obtain that
\begin{equation}
\int\limits_{-H}^{H}f\frac{\partial \omega }{\partial r}dz\approx \frac{%
\partial \overline{\omega }}{\partial r}\int\limits_{-H}^{H}fdz.
\label{VF60}
\end{equation}
We will also use the assumption that the variation of $v_{r}$ with $z$ is
negligible (see, e.g., \cite{Clarke and Carswell 2007}). Then we obtain the
equalities similar to (\ref{VF50}), (\ref{VF60}).
\begin{equation}
\int\limits_{-H}^{H}fv_{r}dz\approx \overline{v}_{r}\int\limits_{-H}^{H}fdz,%
\quad \int\limits_{-H}^{H}f\frac{\partial v_{r}}{\partial r}dz\approx \frac{%
\partial \overline{v}_{r}}{\partial r}\int\limits_{-H}^{H}fdz.  \label{VF65}
\end{equation}

Recall that the dynamic viscosity, $\mu _{v}$, of a gas increases with
temperature, $T$, and, in fact, it is independent of pressure and density at
a given temperature (see, e.g., \cite[p. 186]{Anderson et al. 1984},
\cite[p. 153]{Clarke and Carswell 2007}, \cite[p. 46]{Landau and Lifshitz
1987}, \cite[p. 635]{Loitsyanskiy 1978}, \cite[pp. 26-28]{White 2006}). We
can assume, in view of (\ref{Int54}), (\ref{Int56}), and (\ref{PG35}), that $%
\left. \mu _{v}\right| _{z=\pm H}=0$. Hence,
\begin{equation}
\left. \mu _{v}\frac{\partial v_{\varphi }}{\partial z}\right| _{z=\pm H}=0.
\label{VF90}
\end{equation}

After the averaging of Eq. (\ref{VF07}) over the depth of disk we obtain:
\begin{equation*}
\frac{1}{r^{2}}\overline{\frac{\partial }{\partial r}\left( r^{3}v_{r}\rho
\omega \right) }+\frac{1}{2H}\left[ \left. \left( \rho v_{\varphi
}v_{z}\right) \right| _{z=H}-\left. \left( \rho v_{\varphi }v_{z}\right)
\right| _{z=-H}\right] =
\end{equation*}
\begin{equation}
\frac{1}{r^{2}R_{e}}\overline{\frac{\partial }{\partial r}\left( \mu
_{v}r^{3}\frac{\partial \omega }{\partial r}\right) }+\frac{1}{2HR_{e}}\left[
\left. \left( \mu _{v}\frac{\partial v_{\varphi }}{\partial z}\right)
\right| _{z=H}-\left. \left( \mu _{v}\frac{\partial v_{\varphi }}{\partial z}%
\right) \right| _{z=-H}\right] .  \label{VF100}
\end{equation}
Then, by virtue of (\ref{PG30}), (\ref{VF40})-(\ref{VF60}), (\ref{VF90}),
and (\ref{VF100}), we find that
\begin{equation}
\frac{\partial }{\partial r}\left( 2Hr^{3}\overline{v_{r}\rho }\,\overline{%
\omega }\right) =\frac{1}{R_{e}}\frac{\partial }{\partial r}\left( 2Hr^{3}%
\overline{\mu }_{v}\frac{\partial \overline{\omega }}{\partial r}\right) .
\label{VF110}
\end{equation}
Analogously, we obtain from Eq. (\ref{InvF10}) that
\begin{equation}
\frac{1}{r}\frac{\partial \left( 2Hr\overline{v_{r}\rho }\right) }{\partial r%
}=0,\ \Rightarrow \ 2Hr\overline{v_{r}\rho }\equiv \overline{Q}=const<0.
\label{VF120}
\end{equation}
Let us note that we confine ourselves here to the case when $\overline{Q}%
\neq 0$. By virtue of (\ref{VF120}) and (\ref{VF110}) we obtain the
following steady-state version of the viscous evolution equation (cf.
\cite[p. 166]{Clarke and Carswell 2007}):
\begin{equation}
\frac{\partial }{\partial r}\left( r^{2}\overline{\omega }\overline{Q}%
\right) =\frac{1}{R_{e}}\frac{\partial }{\partial r}\left( r^{3}M_{v}\frac{%
\partial \overline{\omega }}{\partial r}\right) ,\quad M_{v}\equiv 2H%
\overline{\mu }_{v}.  \label{VF130}
\end{equation}

Now we consider the widly used approximation, i.e. that the laminar, yet
viscous disk is nearly Keplerian (see, e.g., \cite{Clarke and Carswell 2007}%
, \cite{Frank et al. 2002}, \cite{Hartmann 2009}, \cite{Vietri Mario 2008}).
Let us use the power-law model \cite[p. 374]{Vietri Mario 2008} for the
angular velocity $\overline{\omega }$, i.e.
\begin{equation}
\overline{\omega }\varpropto \frac{1}{r^{1+\varkappa }},\quad \varkappa
=const,\quad 0.5\leq \varkappa \leq 1.  \label{VF140}
\end{equation}
Let $\hat{M}_{v}\equiv \left. M_{v}\right| _{r=1}$, then, by virtue of (\ref
{VF140}) and (\ref{VF130}), we find:
\begin{equation}
M_{v}=\frac{\hat{M}_{v}}{r^{1-\varkappa }}-\frac{R_{e}\overline{Q}}{%
1+\varkappa }\left( 1-\frac{1}{r^{1-\varkappa }}\right) ,\quad \overline{Q}%
=const<0.  \label{VF160}
\end{equation}
Remark that $M_{v}=const$ if $\varkappa =1$ (the vortex motion). If $%
M_{v}=const$ in (\ref{VF130}), then $\overline{\omega }\varpropto r^{-2}$
fulfills Eq. (\ref{VF130}).

Since we consider the model of laminar viscous disk, it should be mentioned
that the dynamic viscosity, $\mu _{v}$, is often represented as follows: $%
\mu _{v}=\nu \rho $, where $\nu $ denotes the kinematic viscosity (see,
e.g., \cite{Clarke and Carswell 2007}, \cite{Frank et al. 2002}, \cite
{Hartmann 2009}, \cite{Vietri Mario 2008}). It is nothing more than the
definition of the kinematic viscosity, i.e. $\nu \equiv \mu _{v}\diagup \rho
$, (see, e.g, \cite{Clarke and Carswell 2007}, \cite{Goedbloed and Poedts
2004}, \cite{Loitsyanskiy 1978}, \cite{Sedov 1971}, \cite{White 2006}).
Obviously, if the temperature $T=const$, then $\mu _{v}=const$ and, hence, $%
\nu \varpropto \rho ^{-1}$ (in particular, $\nu \rightarrow \infty $ as $%
\rho \rightarrow 0$). Using $\mu _{v}=\nu \rho $, we obtain after the
integrating of Eq. (\ref{VF07}) over the depth of disk (see, e.g., \cite[p.
166]{Clarke and Carswell 2007}) that $M_{v}=\overline{\overline{\nu }}\sum $%
, where $\sum $ denotes the surface density, i.e. $\sum =2H\overline{\rho }$%
. Using the mass-weighted averaging, suggested by A. Favre (see the
references in \cite{Anderson et al. 1984}), i.e.
\begin{equation}
\overline{\overline{f}}\equiv \frac{\overline{f\rho }}{\overline{\rho }},
\label{VF200}
\end{equation}
we obtain
\begin{equation}
\overline{\overline{\nu }}=\frac{\overline{\nu \rho }}{\overline{\rho }}=%
\overline{\left( \frac{\mu _{v}}{\rho }\rho \right) }\diagup \overline{\rho }%
\ \Rightarrow \ \overline{\overline{\nu }}=\frac{\overline{\mu }_{v}}{%
\overline{\rho }}.  \label{VF230}
\end{equation}
In view of (\ref{VF230}), we find that $M_{v}$ $=$ $\overline{\overline{\nu }%
}\sum $ $=$ ($\overline{\mu }_{v}\diagup \overline{\rho }$)$2H\overline{\rho
}$ $=$ $2H\overline{\mu }_{v}$. Hence, the product $\overline{\overline{\nu }%
}\sum $ in \cite[p. 166]{Clarke and Carswell 2007} does not depend on
density (even formally\footnote{%
If the flow will be adiabatic, then $T\varpropto \rho ^{\gamma -1}$. In such
a case we obtain, formally, that $\mu _{v}\varpropto \rho ^{\left( \gamma
-1\right) n}$. For instance, in the case of monatomic gas we can write that $%
\mu _{v}\varpropto \rho ^{0.51}\approx \sqrt{\rho }$, since typically $%
n=0.76 $. However, if $T$ is given, then $\rho $ is given too and vice
versa. Thus, we cannot vary $\rho $ and, hence, $\mu _{v}$\ without varying $%
T$.}), since we consider the thin disk accretion that is highly nonadiabatic
\cite{Shapiro and Teukolsky 2004}. If we assume, for instance, that $%
\overline{\overline{\nu }}=const$ (see, e.g., \cite[Sec. 12.2]{Clarke and
Carswell 2007}), then, in light of (\ref{VF230}), we find that $\overline{%
\mu }_{v}\varpropto \overline{\rho }$. Hence, such an assumption can lead to
wrong conclusions.

Now we can show that a laminar viscous disk tends to be the vortex motion,
i.e. approximately $\overline{\omega }\varpropto r^{-2}$. After the
integration of Eq. (\ref{VF130}) we obtain:
\begin{equation}
\frac{1}{R_{e}}\frac{\partial \overline{\omega }}{\partial r}=\frac{%
\overline{\omega }\overline{Q}}{rM_{v}}-\frac{\overline{c}_{\varphi }%
\overline{Q}}{r^{3}M_{v}},\quad \overline{c}_{\varphi },\overline{Q}=const.
\label{VF235}
\end{equation}
Since $R_{e}\gg 1$, we are looking for the asymptotic expansion to the
solution of (\ref{VF235}) in the following form $\overline{\omega }=\Lambda
\overline{\omega }+\Pi \overline{\omega }$, where $\Lambda \overline{\omega }
$ and $\Pi \overline{\omega }$ are, respectively, the regular and singular
parts of the expansion. Let us assume, at first, that the asymptotic
solution to Eq. (\ref{VF235}) consists only of the regular series. We obtain
in the zeroth approximation:
\begin{equation}
\Lambda _{0}\overline{\omega }=\frac{\overline{c}_{\varphi }}{r^{2}},\quad
\overline{c}_{\varphi }=const.  \label{VF240}
\end{equation}
One can readily see that the regular part, (\ref{VF240}), of zeroth
approximation to the solution of the viscous problem (\ref{VF130}) coincides
with the exact solution of the non-viscous system. Let $\widehat{\overline{%
\overline{v}}}_{r}\equiv \left. \overline{\overline{v}}_{r}\right| _{r=1}$
and let $\widehat{\overline{\omega }}\equiv \left. \overline{\omega }\right|
_{r=1}$. Taking into consideration the singular part of the expansion, we
find that the zeroth approximation, $\overline{\omega }_{0}\equiv \Lambda
_{0}\overline{\omega }+\Pi _{0}\overline{\omega }$, can be written as the
following:
\begin{equation}
\overline{\omega }_{0}=\frac{\overline{c}_{\varphi }}{r^{2}}+\left( \widehat{%
\overline{\omega }}-\overline{c}_{\varphi }\right) \exp \left\{ R_{e}\frac{%
\widehat{\overline{\overline{v}}}_{r}\widehat{\overline{\rho }}}{\widehat{%
\overline{\mu }}_{v}}\left( r-1\right) \right\} ,\quad \widehat{\overline{%
\overline{v}}}_{r}<0.  \label{VF250}
\end{equation}
Thus, in view of (\ref{Int105}) and (\ref{VF250}), we conclude that the
laminar viscous disk can be approximated with a great precision by the
vortex motion (cf., e.g., \cite{Clarke and Carswell 2007}, \cite{Frank et
al. 2002}, \cite{Hartmann 2009}, \cite{Vietri Mario 2008}).

\subsubsection{Turbulent flow\label{VTF}}

Let us estimate the value of $\varkappa $ in the power-law model \cite[p.
374]{Vietri Mario 2008} for the midplane circular velocity $v_{\varphi
}^{\circ }$:

\begin{equation}
v_{\varphi }^{\circ }=\frac{C_{\varphi }^{\circ }}{r^{\varkappa }},\quad
\forall r\in \frak{R}_{a}^{\circ },\quad C_{\varphi }^{\circ },\varkappa
=const,\ 0.5\leq \varkappa \leq 1,  \label{TF05}
\end{equation}
where $\frak{R}_{a}^{\circ }$ $\equiv $ $\left\{ \left( r,z\right) :\right. $
$r\in \left[ 1,\infty \right) $, $\left. z=0\right\} $. The Euler number, $%
E_{u}$, characterizes ``losses'' \cite{Horneber et al. 2012} (the pressure
loss\footnote{%
The Euler number can be written in the following form: $E_{u}=\left( p_{\ast
}-\check{p}\right) \diagup \left( \rho _{\ast }v_{\ast }^{2}\right) $, where
$\check{p}\equiv \left. p\right| _{r\rightarrow \infty }\approx 0$.}
\cite[p. 84]{Kockmann 2013})\ in a flow. We assume that the turbulent disk
should be such that the losses are minimal, i.e. the Euler number should be
as small as possible. Such an approach was already used in \cite{Borisov
2013} to argue that a turbulent disk tends to be Keplerian. In this
subsection the arguments will be considered in more details. Using Prandtl
and Kolmogorov suggestion \cite[p. 230]{Anderson et al. 1984} that the
turbulent viscosity, $\mu _{t}$, is proportional to the square root of the
kinetic energy of turbulence, $\overline{k}$, we evaluate, by means of (\ref
{Int60}), the midplane value of the viscosity $\mu _{t}^{\circ }\equiv
\left. \mu _{t}\right| _{z=0}$ as the following.
\begin{equation}
\mu _{t}^{\circ }=C_{\kappa }L_{\kappa }\rho ^{\circ }\left( \overline{k}%
^{\circ }\right) ^{0.5},\quad \overline{k}^{\circ }\equiv \left. \overline{k}%
\right| _{z=0},\quad C_{\kappa }=const,  \label{TF10}
\end{equation}
where $L_{\kappa }$\ denotes a turbulence length scale, $\rho ^{\circ
}\equiv \left. \rho \right| _{z=0}$.

The variation of the Euler number, $E_{u}$, with the Reynolds number, $R_{e}$%
, obtained with some $\overline{k}$-$\varepsilon $ models, is depicted in
\cite[Fig. 2]{Bhoite and Narasimham 2009}. In the case of laminar flow, as
we can see in \cite[Fig. 2(b)(Problem 2)]{Bhoite and Narasimham 2009}%
\footnote{%
Problem 2 is selected insofar as we consider, in fact, a gas flow around a
spherical object with no-slip boundary conditions at its surface.}, the
smaller $R_{e}$, the larger $E_{u}$ and, hence, for sufficiently low values
of $R_{e}$ the values of $E_{u}$ are greater than that for turbulent flow.
In such a case we assume that all turbulent terms are eliminated in (\ref
{Int140}), more over, it is assumed that the midplane velocity $v_{r}^{\circ
}\equiv \left. v_{r}\right| _{z=0}=0$ in (\ref{Int140}). Let us remind (see
Sec. \ref{Average equations}) that a laminar viscous disk tends to be the
vortex motion provided that $v_{r}^{\circ }\neq 0$. We will also use the
following asymptotic expansion
\begin{equation}
T^{\circ }\left( r\right) \equiv \left. T\right| _{z=0}\approx
\sum\limits_{i=0}^{\infty }\theta _{i}r^{-i},\quad \theta _{i}=const,
\label{NMNV750}
\end{equation}
in the limit $r\rightarrow \infty $. As usually \cite{Jones 1997}, the
series in (\ref{NMNV750}) may converge or diverge, but its partial sums are
good approximations to $T^{\circ }\left( r\right) $ for large enough $r$.
Assuming that $T^{\circ }\left( r\right) \rightarrow 0$ as $r\rightarrow
\infty $, we find that $\theta _{0}=0$. Thus, we can use the following
approximation for the disk temperature profile:
\begin{equation}
T^{\circ }\left( r\right) =\frac{\theta _{1}}{r}+O\left( r^{-2}\right)
,\quad \theta _{1}=const.  \label{TF25}
\end{equation}
It is also assumed that the characteristic quantity, $T_{\ast }$, for
temperature is taken such that $\theta _{1}=1$. Using (\ref{Int190}), (\ref
{Int200}), and the above assumptions, we obtain the following equation,
which is the steady-state version of Eq. (\ref{Int140}) in the midplane
variables:

\begin{equation}
\frac{\rho ^{\circ }\left( v_{\varphi }^{\circ }\right) ^{2}}{r}=E_{u}\frac{%
\partial P^{\circ }}{\partial r}+\frac{1}{F_{r}}\frac{\rho ^{\circ }}{r^{2}}%
,\quad r>1.  \label{TF27}
\end{equation}
In view of (\ref{TF25}), we have, in the first approximation, that the
midplane temperature $T^{\circ }\left( r\right) =r^{-1}$. In view of (\ref
{Int175}), we write $P^{\circ }=\rho ^{\circ }T^{\circ }$. Then, by virtue
of (\ref{TF05}), we obtain:
\begin{equation}
\rho ^{\circ }=\frac{\hat{\rho}^{\circ }}{r^{\alpha -1}}\exp \zeta \left[
\frac{1}{r^{2\varkappa -1}}-1\right] ,\quad \zeta =\frac{\left( C_{\varphi
}^{\circ }\right) ^{2}}{E_{u}\left( 2\varkappa -1\right) },\ \alpha =\frac{1%
}{E_{u}F_{r}},  \label{TF30}
\end{equation}
where $\hat{\rho}^{\circ }\equiv \left. \rho ^{\circ }\right| _{r=1}$ and $%
0.5<\varkappa \leq 1.$

We obtain from (\ref{TF10}), by virtue of (\ref{TF30}), that
\begin{equation}
\mu _{t}^{\circ }=\frac{C_{\rho }}{r^{\alpha -1}}\exp \zeta \left[ \frac{1}{%
r^{2\varkappa -1}}-1\right] ,\quad C_{\rho }\equiv \hat{\rho}^{\circ
}C_{\kappa }L_{\kappa }\left( \overline{k}^{\circ }\right) ^{0.5}.
\label{TF40}
\end{equation}
If $v_{r}=0$, then the steady-state version of Eq. (\ref{Int150}) can be
written in the following form (see, e.g., \cite{Regev 1983}), provided that $%
\mu _{v}\ll \mu _{t}$.
\begin{equation}
\frac{\partial }{\partial r}\left[ \mu _{t}^{\circ }r^{3}\frac{\partial }{%
\partial r}\left( \frac{v_{\varphi }^{\circ }}{r}\right) \right] =0.
\label{TF50}
\end{equation}
Let us find $\mu _{t}^{\circ }=\mu _{t}^{\circ }\left( r\right) $ such that
Eq. (\ref{TF50}) will be fulfilled by the power-law model (\ref{TF05}). By
virtue of (\ref{TF05}), we find from (\ref{TF50}) that
\begin{equation}
\mu _{t}^{\circ }=\frac{A_{\mu }}{r^{1-\varkappa }},\quad A_{\mu }=const.
\label{TF60}
\end{equation}
Equating (\ref{TF40}) and (\ref{TF60}) at $r=1$, we find that $\left.
C_{\rho }\right| _{r=1}=A_{\mu }$. Let us assume, for the sake of simplicity%
\footnote{%
We may assume (e.g., \cite{Shakura and Sunyaev 1973}) that $\overline{k}%
\propto c_{s}^{2}$. Since $c_{s}^{2}=\partial P/\partial \rho $, we find, in
view of (\ref{Int175}) and (\ref{TF25}), that $\overline{k}^{\circ }\propto
1/r$. Hence, we will get a little more cumbersome formulas, but not
fundamental difficulties.}, that $\overline{k}^{\circ }\approx \widehat{%
\overline{k}}^{\circ }\equiv \left. \overline{k}^{\circ }\right| _{r=1}$ in
the $\varepsilon $-vicinity ($\varepsilon \ll 1$) of $r=1$, i.e. at $r\in
\left( 1,1+\varepsilon \right) $. Let us now assume that $\mu _{t}^{\circ }$
of (\ref{TF40}) and $\mu _{t}^{\circ }$ of (\ref{TF60}) (having, in general,
different values) coincide each other in the vicinity of $r=1$ with accuracy
$O\left( \varepsilon ^{2}\right) $. In such a case we obtain that
\begin{equation}
E_{u}\propto \frac{1}{2-\varkappa }.  \label{TF70}
\end{equation}
As we can see from (\ref{TF70}) and (\ref{TF05}), $E_{u}\rightarrow \min $
if $\varkappa \rightarrow 0.5$. Thus, in the frame of our assumptions, we
find that the turbulent flow tends to be Keplerian. Based on this
conclusion, we can estimate the turbulent viscosity, $\mu _{t}$, provided
that the dynamic viscosity $\mu _{v}\ll \mu _{t}$.

From the above discussion it follows that the magnitude of the centrifugal
force is approximately equal to the gravitational force in the case of
turbulent disk. Hence, we assume for the case of turbulent flow:
\begin{equation}
\frac{\rho v_{\varphi }^{2}}{r}=\frac{\rho }{F_{r}}\frac{\partial \Phi }{%
\partial r}\ \Rightarrow \ \frac{v_{\varphi }^{2}}{r}=\frac{1}{F_{r}}\left(
\frac{1}{r^{2}}-\frac{3}{2r^{4}}z^{2}+\ldots \right) .  \label{TF80}
\end{equation}
Taking into account (\ref{Int200}), i.e. $v_{\varphi }$ $=$ $v_{\varphi
}^{\circ }$ $+$ $v_{\varphi }^{\prime \prime \circ }z^{2}$ $+$ $\ldots $, we
obtain from (\ref{TF80}) that
\begin{equation}
v_{\varphi }^{\circ }=\frac{1}{\sqrt{rF_{r}}},\quad v_{\varphi }^{\prime
\prime \circ }=-\frac{3v_{\varphi }^{\circ }}{4r^{2}}\text{.}  \label{TF90}
\end{equation}
After simple transformations, the steady-state version of Eq. (\ref{Int150}%
), where $\mu =\mu _{v}+\mu _{t}$, can be written in the following form.
\begin{equation}
\frac{\rho v_{r}}{r}\frac{\partial }{\partial r}\left( rv_{\varphi }\right)
+\rho v_{z}\frac{\partial v_{\varphi }}{\partial z}=\frac{1}{r^{2}R_{e}}%
\frac{\partial }{\partial r}\left( \mu _{t}r^{3}\frac{\partial \omega }{%
\partial r}\right) +\frac{1}{R_{e}}\frac{\partial }{\partial z}\left( \mu
_{t}\frac{\partial v_{\varphi }}{\partial z}\right) ,  \label{TF100}
\end{equation}
where $\omega $ denotes the angular velocity. By virtue of (\ref{Int190}), (%
\ref{Int200}), and (\ref{TF90}), we can write the momentum equation (\ref
{TF100}) in the midplane variables, i.e. at $z=0$, as the following:
\begin{equation*}
\rho ^{\circ }v_{r}^{\circ }\frac{\partial }{\partial r}\left( r^{2}\omega
^{\circ }\right) =
\end{equation*}
\begin{equation}
\frac{1}{rR_{e}}\frac{\partial }{\partial r}\left( \mu _{t}^{\circ }r^{3}%
\frac{\partial \omega ^{\circ }}{\partial r}\right) -\frac{3\omega ^{\circ }%
}{2R_{e}}\mu _{t}^{\circ },\quad \omega ^{\circ }\equiv \left. \omega
\right| _{z=0},\ \mu _{t}^{\circ }\equiv \left. \mu _{t}\right| _{z=0}.
\label{TF110}
\end{equation}
Since the disk tends to be Keplerian, i.e. $\omega ^{\circ }\varpropto
r^{-1.5}$, we obtain from (\ref{TF110}) the following equation in the
unknown $\mu _{t}^{\circ }$.
\begin{equation}
\frac{\partial \mu _{t}^{\circ }}{\partial r}+\frac{3\mu _{t}^{\circ }}{2r}+%
\frac{R_{e}\rho ^{\circ }v_{r}^{\circ }}{3}=0.  \label{TF120}
\end{equation}

Let $z=h\left( r\right) $ denote a streamline, namely a line that is tangent
to the meridional velocity vector, ($v_{r},v_{z}$), and let $\hat{h}\equiv
\left. h\right| _{r=1}$. We assume that $\mid \hat{h}\mid \ll 1$ and $\hat{h}%
\left( r\right) \equiv 0$ if $\hat{h}=0$, i.e. the streamline $z=h\left(
r\right) $ is neighboring to the streamline $z=0$. It is also assumed that $%
\hat{h}\ncong 0$. Consequently, the kinematic condition (see, e.g., \cite[p.
165]{Sedov 1971}, \cite[p. 50]{White 2006}) at the streamline will be the
following.
\begin{equation}
\frac{d}{dt}\left( z-h\right) =0\Rightarrow \left. v_{z}\right| _{z=h}=\frac{%
\partial h}{\partial r}\left. v_{r}\right| _{z=h}.  \label{CVB20}
\end{equation}
Using (\ref{Int200}) and the following transformation on $h$:
\begin{equation}
h\rightarrow \hat{h}h,  \label{CVB30}
\end{equation}
we rewrite (\ref{CVB20}) to read:
\begin{equation}
v_{z}^{\prime \circ }\hat{h}h+v_{z}^{\prime \prime \prime \circ }\hat{h}%
^{3}h^{3}+\ldots =\frac{\hat{h}\partial h}{\partial r}\left( v_{r}^{\circ
}+v_{r}^{\prime \prime \circ }\hat{h}^{2}h^{2}+\ldots \right) .
\label{CVB40}
\end{equation}
From this equation we obtain with the accuracy $O$($\hat{h}^{2}$), that
\begin{equation}
v_{z}^{\prime \circ }=v_{r}^{\circ }\frac{\partial \ln h}{\partial r}.
\label{CVB50}
\end{equation}
We remark that (\ref{CVB50}) is invariant for the inverse transformation on $%
h$, i.e. $h\leftarrow \hat{h}h$. Let us consider the case when $h$ is a
linear function of $r$, i.e.
\begin{equation}
h=\hat{h}r\ \Rightarrow \ v_{z}^{\prime \circ }=\frac{v_{r}^{\circ }}{r}.
\label{CVB60}
\end{equation}
In view of (\ref{CVB60}), we\ rewrite (\ref{InvF80}) in the following form:
\begin{equation}
\frac{\partial \left( \rho ^{\circ }v_{r}^{\circ }\right) }{\partial r}+%
\frac{2}{r}\left( \rho ^{\circ }v_{r}^{\circ }\right) =0.  \label{CVB70}
\end{equation}
Let us note that the equations (\ref{TF120}) and (\ref{CVB70}) are linear in
$\mu _{t}^{\circ }$ and $\left( \rho ^{\circ }v_{r}^{\circ }\right) $. We
assume, that
\begin{equation}
\mu _{t}^{\circ }=a\left( \rho ^{\circ }v_{r}^{\circ }\right) +b,\quad
a=a\left( r\right) ,\ b=b\left( r\right) .  \label{CVB80}
\end{equation}
In such a case (\ref{TF120}) can be written as follows:
\begin{equation}
\frac{\partial \left( \rho ^{\circ }v_{r}^{\circ }\right) }{\partial r}%
+\left( \frac{\partial a}{\partial r}+\frac{3}{2r}+\frac{R_{e}}{3a}\right)
\left( \rho ^{\circ }v_{r}^{\circ }\right) +\frac{\partial b}{a\partial r}+%
\frac{3b}{2ar}=0.  \label{CVB90}
\end{equation}
From (\ref{CVB70}) and (\ref{CVB90}) we obtain the following equations for $%
a\left( r\right) $ and $b\left( r\right) $.
\begin{equation}
\frac{\partial a}{\partial r}+\frac{3}{2r}+\frac{R_{e}}{3a}=\frac{2}{r},
\label{CVB100}
\end{equation}
\begin{equation}
\frac{\partial b}{a\partial r}+\frac{3b}{2ar}=0.  \label{CVB110}
\end{equation}
Then, we find from (\ref{CVB100}), (\ref{CVB110}) that
\begin{equation}
a=-\frac{2R_{e}}{3}\left( r+C_{a}^{\circ }\sqrt{r}\right) ,\quad
b=C_{b}^{\circ }r^{-1.5},\quad C_{a}^{\circ },\ C_{b}^{\circ }=const.
\label{CVB120}
\end{equation}
Thus, in view of (\ref{CVB80}) and (\ref{CVB120}), we find:
\begin{equation}
\mu _{t}^{\circ }=R_{e}\left[ -\frac{2}{3}\rho ^{\circ }v_{r}^{\circ
}(r+C_{a}^{\circ }\sqrt{r})\right] +C_{b}^{\circ }r^{-1.5},\ v_{r}^{\circ
}\leq 0.  \label{TF130}
\end{equation}
Following Prandtl and Kolmogorov (e.g., \cite[p. 230]{Anderson et al. 1984},
\cite[p. 74]{Wilcox 1994}) we expect that the turbulent viscocity, $\mu
_{t}^{\circ }$, can be modeled as (\ref{TF10}). In view (\ref{TF10}), the
turbulent viscocity is proportional to $\rho ^{\circ }$ and, hence,\ we may
assume that $C_{b}^{\circ }=0$. Assuming that the disk semi-thickness $%
H\varpropto r+C_{a}^{\circ }\sqrt{r}$, we obtain from (\ref{TF130}):
\begin{equation}
\mu _{t}^{\circ }=R_{e}b_{\mu }^{\circ }\rho ^{\circ }\left( \overline{k}%
^{\circ }\right) ^{0.5}H,\quad b_{\mu }^{\circ }=\frac{2\left(
1+C_{a}^{\circ }\right) }{3\hat{H}},\ H=\hat{H}\frac{r+C_{a}^{\circ }\sqrt{r}%
}{1+C_{a}^{\circ }},  \label{TF135}
\end{equation}
where $\hat{H}=\left. H\right| _{r=1}$, and
\begin{equation}
\overline{k}^{\circ }\sim \left( v_{r}^{\circ }\right) ^{2},\quad R_{e}\gg 1.
\label{TF140}
\end{equation}
We can conclude from (\ref{TF135}) and (\ref{Int120}) that $\mu _{t\ast
}\diagup \mu _{v\ast }$ $\sim $ $R_{e}$ (see also, e.g., \cite[p. 213]
{Belotserkovskii et al. 2005}). It is necessary to stress that the
expression (\ref{TF140}) for $\overline{k}^{\circ }$ is nothing more than
the estimation of the turbulent kinetic energy provided that all of the
above assumptions are valid. The more correct approach is to use the
one-equation model (\ref{Int70}) to estimate the turbulent kinetic energy, $%
\overline{k}^{\circ }$, in (\ref{TF135}).

Let us note, if we assume that the turbulent viscosity can be represented in
the form (\ref{TF135}) then, using the following transformation on $\mu
_{t}^{\circ }$%
\begin{equation}
\mu _{t}^{\circ }\longrightarrow R_{e}\mu _{t}^{\circ }  \label{TF150}
\end{equation}
we find that the PDEs are not singularly perturbed, namely, the small
parameter, $1/R_{e}$, does not premultiply the highest-order
turbulent-viscous terms.

\section{Conclusion}

By and large axisymmetric steady-state perfect gas flows are investigated.
First of all, it is considered the case of inviscid flow (Sec. \ref{Circular
velocity}) with a non-zero radial velocity, $v_{r}\neq 0$. It is shown that
the only exact solution for the circular velocity, $v_{\varphi }$, can be a
vortex. It is important that $v_{\varphi }$ may differ from the vortex only
if $v_{r}=0$. In the case $v_{r}=0$, it is used the popular power-law model
for the temperature distribution to demonstrate that the midplane circular
velocity can differ significantly from the Keplerian one. Special attention
is devoted to the ``proof'' that accretion disks are Keplerian in zeroth
approximation. The ``proof'' was done in many works by the regular
perturbation technique, applied to singularly perturbed PDEs. It is
demonstrated that such an approach can lead to erroneous conclusion and,
hence, the singular part of the asymptotic expansion is usually necessary
for correct solution of the singularly perturbed problem. Thus, the
well-known arguments that an inviscid accretion disk must be sub-Keplerian
are disproved.

The instability of inviscid non-vortex flow investigated in Sec. \ref
{Instability of inviscid disk}. It is considered the steady-state accretion
disk as input-output system. The boundary conditions are assumed as inputs.
The mapping $F$: input $\rightarrow $ output is defined by the steady-state
balance equations as well as by the equation of state. It is shown that an
infinitesimal increment in the input triggers the finite increment in the
output, which is to say that the flow is unstable. After that, the stability
of the regular part of asymptotic solution to the singularly perturbed
system is examined. It is found that the infinitesimal increment in the
input does not trigger a finite increment in the output and, hence, the
motion in question is stable. Thus, the assumption that the asymptotic
solution to the singularly perturbed PDEs consists only of the regular
series can lead to a wrong conclusion about the stability of the system.

A laminar viscous disk dynamics is investigated in Sec. \ref{Average
equations}. The accretion disk is assumed geometrically thin. It is used the
well known approach for the investigation of the accretion disk dynamics,
namely, the equations of mass and momentum conservation are averaged over
the depth of disk. It is assumed that the asymptotic solution consists of
the regular as well as singular series The found zeroth approximation shows
that the laminar viscous disk tends to be the vortex motion. Thus, the
well-known arguments that a laminar viscous accretion disk must be
sub-Keplerian are questionable.

Turbulent flow is studied in Sec. \ref{VTF}. It is found that the turbulent
flow tends to be Keplerian. This result was obtained on the basis that a
turbulent gas tends to flow with minimal losses, i.e., the Euler number
should be as small as possible. Based on the fact that the midplane circular
velocity is Keplerian, the turbulent viscosity is evaluated. It is
demonstrated that turbulent viscosity, $\mu _{t}$, is proportional to the
Reynolds number, $R_{e}$. Hence, using the obvious transformation on the
turbulent viscosity, the non-singularly perturbed PDEs can be obtained,
namely, the small parameter, $1/R_{e}$, will not premultiply the
highest-order turbulent-viscous terms.


\bigskip

\begin{thebibliography}{999}
\bibitem{Anderson et al. 1984}  D. A. Anderson, J. C. Tannehill, and R. H.
Pletcher, Computational fluid mechanics and heat transfer, Hemisphere
Publishing Corporation, New York, 1984.

\bibitem{Archambaul 2002}  Mark R. Archambault, Oshin Peroomian,
Characterization of a gas/gas, hydroge/oxygen engine, AIAA 2002-3594, 38th
AIAA/ASME/SAE/ASEE Joint Propulsion Conference \& Exhibit 7-10 July 2002,
Indianapolis, Indiana

\bibitem{Armijo 2012}  Matias Montesinos Armijo, Review: Accretion disk
theory, arXiv:1203.6851 [astro-ph.HE], 2012

\bibitem{Armitage 2010}  P. Armitage, Astrophysics of Planet Formation,
Cambridge University Press, UK, 2010.

\bibitem{Arnol'd 1992}  Vladimir I. Arnol'd, Ordinary Differential
Equations, Springer-Verlag Berlin Heidelberg ,1992.

\bibitem{Belotserkovskii et al. 2005}  O. M. Belotserkovskii, A. M. Oparin,
V. M. Chechetkin, Turbulence: New Approaches, Cambridge Internationl Science
Publishing Ltd, 2005

\bibitem{Beskin et al 2002}  V. Beskin, G. Henri, F. M\'{e}nard, G.
Pelletier, and J. Dalibard, eds. Accretion discs, jets and high energy
phenomena in astrophysics, Springer-Verlag, Berlin, 2003

\bibitem{Bhoite and Narasimham 2009}  Mayur T. Bhoite, G.S.V.L. Narasimham,
Turbulent mixed convection in a shallow enclosure with a series of heat
generating components, International Journal of Thermal Sciences 48 (2009)
948-963

\bibitem{Bisnovatyi-Kogan 2004}  Gennady Bisnovatyi-Kogan, Accretion disks
around black holes with account of magnetic fields, arXiv:astro-ph/0406466,
June 2004.

\bibitem{Blackburne 2011}  Blackburne Jeffrey A., Pooley David, Rappaport
Saul, and Schechter Paul L., Sizes and Temperature Profiles of Quasar
Accretion Disks from Chromatic Microlensing. The Astrophysical Journal 729.1
(2011): 34., see also in arXiv:1007.1665v2 [astro-ph.CO], 2011

\bibitem{Borisov 2013}  V. S. Borisov, On dynamics of geometrically thin
accretion disks, arXiv:1304.7459v5 [astro-ph.SR], 2013

\bibitem{Boss 2005}  A. P. Boss, The Solar Nebula, in the first volume of
Treatise on Geochemistry. Meteorites, Comets and Planets, volume editor
Andrew M. Davis, Elsevier Ltd., Oxford, UK, 2005.

\bibitem{Chaudhuri 2016}  Chaudhuri Anuja Ray, A Comparison Study of
Non-Standard Analysis and Non-Archimedean Ultrametric Theory, International
Journal of Emerging Trends in Science and Technology, Vol. 03 Issue 02,
Pages 3601-3607, 2016.

\bibitem{Clarke and Carswell 2007}  Cathie Clarke and Bob Carswell,
Principles of Astrophysical Fluid Dynamics, Cambridge University Press, New
York, 2007.

\bibitem{Doetsch 1961}  G. Doetsch, Guide to the applications of Laplace
transforms, D. Van Nostrad Company, London, 1961.

\bibitem{Duric 2004}  Neb Duric, Advanced Astrophysics, Cambridge University
Press, New York, 2004.

\bibitem{Flett 2008}  Flett T. M., Differential analysis, Cambridge
University Press, New York, 2008.

\bibitem{Frank et al. 2002}  Juhan Frank, Andrew King, and Derek Raine,
Accretion Power in Astrophysics. Cambridge University Press, New York, 2002.

\bibitem{Frick 2003}  P. G. Frick, Turbulence: Approaches and Models, IKI,
Moscow-Izhevsk, 2003 (in Russian).

\bibitem{Fridman et al 2006}  Aleksey M. Fridman (ed.), M.Y. Marov (ed.),
and Ilya G. Kovalenko (ed.), Astrophysical Disks: Collective and Stochastic
Phenomena, Springer, Dordrecht, The Netherlands, 2006.

\bibitem{Garcia 2011}  Paulo J. V. Garcia (ed.), Physical processes in
circumstellar disks around young stars, The University of Chicago Press,
USA, 2011.

\bibitem{Glaz 1981}  H. M. Glaz, Statistical behavior and coherent
structures in two dimensional inviscid turbulence, SIAM J. Appl. Math., 41,
459-479, (1981).

\bibitem{Goedbloed and Poedts 2004}  J. P. Goedbloed and S. Poedts,
Principles of magnetohydrodynamics: with applications to laboratory and
astrophysical plasmas, Cambridge University Press, Cambridge, 2004.

\bibitem{Harris 1983}  C. J. Harris, The stability of input-utput dynamical
system, Academic Press, London, 1983.

\bibitem{Harsono et al. 2013}  D. Harsono, R. Visser, S. Bruderer, E. F. van
Dishoeck, and L. E. Kristensen, Evolution of CO lines in time-dependent
models of protostellar disk formation, A\&A 555, A45 (17pp), 2013.

\bibitem{Hartmann 2009}  Lee Hartmann, Accretion processes in star
formation, Cambridge University Press, Cambridge, 2009.

\bibitem{Hirschel 2005}  Ernst Heinrich Hirschel, Basics of
aerothermodynamics, Springer, Berlin, 2005.

\bibitem{Horneber et al. 2012}  T. Horneber, C. Rauh, and A. Delgado, Fluid
dynamic characterisation of porous solids in catalytic fixed-bed reactors,
Microporous and Mesoporous Materials 154 (2012) 170-174

\bibitem{Hure et al. 2011}  J.-M. Hure, F. Hersant, C. Surville, N. Nakai,
and T. Jacq, AGN disks and black holes on the weighting scales, A\&A 530,
A145 (8 pp), 2011.

\bibitem{Il'in 1992}  A. M. Il'in, Matching of asymptotic expansions of
solutions of boundary value problems, American Mathematical Society, USA,
1992.

\bibitem{Jimenez-Vicente 2014}  J. Jimenez-Vicente, E. Mediavilla, C. S.
Kochanek, J. A. Munoz, V. Motta, E. Falco, and A. M. Mosquera, The Average
Size and Temperature Profile of Quasar Accretion Disks, The Astrophysical
Journal, 783:47 (7pp), 2014 March 1.

\bibitem{Jin-Lu Yu 2013}  Jin-Lu Yu, Li-ming He, Yi-fei Zhu ,Wei Ding,
Yu-qian Wang, Numerical simulation of the effect of plasma aerodynamic
actuation on improving film hole cooling performance, Heat Mass Transfer 49,
897-906, 2013.

\bibitem{Jones 1997}  D. S. Jones, Introduction to Asymptotics: A Treatment
Using Nonstandard Analysis, World Scientific Publishing, London, 1997.

\bibitem{Kawamura 2010}  Kiko Kawamura, The derivative of\ Lebesgues
singular function, Real Analysis Exchange, Summer Symposium 2010, pp. 83-85

\bibitem{Kawamura 2011}  Kiko Kawamura, On the set of points where Lebesgues
singular function has the derivative zero, Proc. Japan Acad., 87, Ser. A,
162-166, 2011

\bibitem{Keisler 2012}  Keisler H. Jerome, Elementary Calculus: An
Infinitesimal Approach. On-line Edition, http://www.math.wisc.edu/\symbol{126%
}keisler/calc.html. Copyright \copyright\ 2000 by H. Jerome Keisler, revised
February 2012.

\bibitem{Kockmann 2013}  N. Kockmann (ed.), Micro Process Engineering,
Wiley-VCH Verlag GmbH \& Co. KGaA, Weinheim, Germany, 2013.

\bibitem{Kolmogorov and Fomin 1970}  A. N. Kolmogorov and S. V. Fomin,
Introductory Real Analysis, Prentice-Hall, Inc., Englewood Cliffs, USA, 1970.

\bibitem{Kluzniak and Kita 2000}  W. Kluzniak, D. Kita, Three-dimensional
structure of an alpha accretion disk, arXiv:astro-ph/0006266, 2000

\bibitem{Kudritzki and Puls 2000}  Rolf-Peter Kudritzki and Joachim Puls,
Winds from Hot Stars, Annu. Rev. Astron. Astrophys. 2000. 38:613-66

\bibitem{Kurbatov et al. 2014}  E. P. Kurbatov, D. V. Bisikalo, and P. V.
Kaygorodov, On the possible turbulence mechanism in accretion disks in
non-magnetic binary stars, Physics - Uspekhi 57 (8) 787 - 798 (2014).

\bibitem{Kuznetsov 2005}  O. A. Kuznetsov, On the excitation of
hydrodynamical turbulence in accretion discs, AIP Conference Proceedings
797, 271 (2005); doi: 10.1063/1.2130243.

\bibitem{Kuznetsov 2006}  O. A. Kuznetsov, Hydrodynamical Turbulence in
Accretion Discs, in Astrophysical Disks: Collective and Stochastic
Phenomena, volume editors: Aleksey M. Fridman, M.Y. Marov, and Ilya G.
Kovalenko, Springer, Dordrecht, The Netherlands, 2006.

\bibitem{Lai et al. 2010}  W. Michael Lai, David Rubin, and Erhard Krempl,
Introduction to Continuum Mechanics, Elsevier Inc., Oxford, UK, 2010

\bibitem{Lakshmikantham 2015}  V. Lakshmikantham, S. Leela, A. A. Martynyuk,
Stability Analysis of Nonlinear Systems, Springer International Publishing
Switzerland, 2015.

\bibitem{Landau and Lifshitz 1987}  L. D. Landau and E. M. Lifshitz, Fluid
mechanics, Pergamon Books Ltd., Oxford, 1987.

\bibitem{Liverts et al. 2012}  E. Liverts, Yu. Shtemler and M. Mond, Linear
and Weakly Nonlinear Analysis of the Magneto-Rotational-Instability in Thin
Keplerian Discs, arXiv:1201.2847v1 [astro-ph.SR] 13 Jen 2012.

\bibitem{Lurie and Enright 2012}  Boris J. Lurie and Paul J. Enright,
Classical Feedback Control: With MATLAB\registered\ and Simulink\registered
, Taylor \& Francis Group, LLC, Boca Raton, FL, USA, 2012.

\bibitem{Edward Liverts et al. 2012}  Edward Liverts, Yuri Shtemler, Michael
Mond, Orkan M Umurhan, and Dmitry V Bisikalo, Non-Dissipative Saturation of
the Magnetorotational Instability in Thin Disks, arXiv:1210.5343v1
[astro-ph.SR] 19 Oct 2012.

\bibitem{Loitsyanskiy 1978}  L.G. Loitsyanskiy, \textit{Mechanics of Liquids
and Gases}, Nauka, Moscow, 1978 (in Russian)

\bibitem{Malanchev et al. 2019}  Konstantin L. Malanchev, Konstantin A.
Postnov, Nikolay I. Shakura, Physical conditions in thin laminar-convective
accretion flows, 4th International Conference on Particle Physics and
Astrophysics (ICPPA-2018), Journal of Physics: Conference Series 1390 (2019)
012085.

\bibitem{Marcus et al. 2015}  Philip S. Marcus, Suyang Pei, Chung-Hsiang
Jiang, Joseph A. Barranco, Pedram Hassanzadeh, and Daniel Lecoanet, Zombie
vortex instability. I. A purely hydrodynamic instability to resurrect the
dead zones of protoplanetary disks, The Astrophysical Journal, 808:87
(16pp), 2015 July 20.

\bibitem{Mesarovic and Takahara 1975}  M. D. Mesarovic and Yasuhiko
Takahara, General Systems Theory: Mathematical Foundations, Academic Press,
New York, 1975.

\bibitem{Min-Kai Lin and Youdin 2015}  Min-Kai Lin and Andrew N Youdin,
Cooling Requirements for the Vertical Shear Instability in Protoplanetary
Disks, arXiv:1505.02163v2 [astro-ph.EP], 14 Jul 2015

\bibitem{Mishchenko and Rozov  1980}  E. F. Mishchenko and N. Kh. Rozov,
Differential equations with small parameters and relaxation oscillations,
Plenum Press, New York, 1980.

\bibitem{Moiseev 1981}  N. N. Moiseev, Mathematical Problems of System
Analysis. Nauka, Moscow, 1981 (in Russian).

\bibitem{Mukhopadhyay and Chattopadhyay 2013}  Banibrata Mukhopadhyay and
Amit K Chattopadhyay, Stochastically driven instability in rotating shear
flows, J. Phys. A: Math. Theor. 46 035501 (17pp), 2013

\bibitem{Murillo et al. 2013}  Nadia M. Murillo, Shih-Ping Lai, Simon
Bruderer, Daniel Harsono, and Ewine F. van Dishoeck, A Keplerian disk around
a Class 0 source: ALMA observations of VLA1623A, A\&A 560, A103 (16 pp),
2013.

\bibitem{Nature Publishing Group 2001}  Paul Murdin (ed.), Encyclopedia of
Astronomy and Astrophysics, Nature Pub. Group, Institute of Physics Pub.,
London, Bristol, 2001.

\bibitem{Nelson et al 2013}  Richard P. Nelson, Oliver Gresse, and Orkan M.
Umurhan, Linear and non-linear evolution of the vertical shear instability
in accretion discs, MNRAS 435, 2610-2632, 2013.

\bibitem{Ogilvie 1997}  G. I. Ogilvie, The equilibrium of a differentially
rotating disc containing a poloidal magnetic field, Mon. Not. R. Astron.
Soc. 288, 63-77, 1997.

\bibitem{Parker 1974}  D. A. Parker, The equilibrium of an interstellar
magnetic disk, Mon. Not. R. astr. Soc., 168, 331-344, 1974.

\bibitem{Piran 1978}  Tzvi Piran, The role of viscosity and cooling
mechanisms in the stability of accretion disks, The Astrophysical Journal,
221: 652-660, 1978.

\bibitem{Pittard et al. 2009}  J. M. Pittard, S. A. E. G. Falle, T. W.
Hartquist, and J. E. Dyson, The turbulent destruction of clouds I. A k-e
treatment of turbulence in 2D models of adiabatic shockcloud interactions,
Mon. Not. R. Astron. Soc. 394, 1351-1378 (2009)

\bibitem{Ponstein 2001}  J. Ponstein, Nonstandard Analysis, ISBN:
90-367-1672-1.

\bibitem{Pope 2000}  Stephen B. Pope, Turbulent flows, Cambridge University
Press, New York, 2000.

\bibitem{Pringle 1981}  J. E. Pringle, 'Accretion discs in astrophysics,
Ann. Rev. Astron. Astrophys. 19:137-62,1981.

\bibitem{Raettig et al. 2013}  Natalie Raettig, Wladimir Lyra, Hubert Klahr,
A Parameter Study for Baroclinic Vortex Amplification, The Astrophysical
Journal, 765:115 (12pp), 2013.

\bibitem{Razdoburdin 2020}  D.N. Razdoburdin, Perturbations dynamics in
Keplerian flow under external stochastic forcing, arXiv:2001.03912v1
[astro-ph.HE], 12 Jen 2020.

\bibitem{Razdoburdin and  Zhuravlev 2015}  D.N. Razdoburdin, V.V. Zhuravlev,
Transient dynamics of perturbations in astrophysical disks,
arXiv:1512.08897v1 [astro-ph.HE], 30 Dec 2015.

\bibitem{Reddy 2008}  J. N. Reddy, An Introduction to Continuum Mechanics,
with Applications, Cambridge University Press, Cambridge, UK, 2008.

\bibitem{Rebusco et al 2009}  P. Rebusco, O.M. Umurhan, W. Klu%
\'{}%
zniak, and O. Regev, Global transient dynamics of three-dimensional
hydrodynamical disturbances in a thin viscous accretion disk, Physics of
fluids 21, 076601 2009 (arXiv:0906.0004v2 [astro-ph.HE] 11 Jun 2009)

\bibitem{Regev 1983}  O. Regev, The disk-star boundary layer and its effect
on the accretion disk structure, Astron. and Astrophys. vol. 126, no. 1,
146-151, 1983.

\bibitem{Regev 2008}  Oded Regev, Hydrodynamical activity in thin accretion
disks, New Astronomy Reviews 51 (2008) 819-827

\bibitem{Regev et al. 2016}  Oded Regev, Orkan M. Umurhan, Philip A. Yecko,
Modern Fluid Dynamics for Physics and Astrophysics, Springer, New York, 2016.

\bibitem{Samoylenko 2004}  Yuliya Samoylenko, Asymptotical Expansions for
One-Phase Soliton-Type Solution to Perturbed Korteweg -- de Vries Equation,
Proceedings of Institute of Mathematics of NAS of Ukraine, 2004, Vol. 50,
Part 3, 1435-1441

\bibitem{Sanchez 2014}  Francisco Sanchez (ed), Accretion Processes in
Astrophysics, Cambridge University Press 2014.

\bibitem{Schlichting 1968}  Schlichting Herrmann, Boundary-Layer Theory, 6th
Edition, McGraw-Hill, New York, 1968.

\bibitem{Schade and Neemann 2018}  Heinz Schade, Klaus Neemann, Tensor
Analysis, Walter de Gruyter GmbH, Berlin/Boston, 2018.

\bibitem{Schroder 2008}  Bernd S. W. Schr\"{o}der, Mathematical Analysis, A
Concise Introduction, John Wiley \& Sons, Inc., Hoboken, New Jersey, 2008.

\bibitem{Sedov 1971}  L. I. Sedov, A course in continuum mechanics\textit{, }
Wolters-Noordhoff Publishing, Groningen, the Netherlands, 1971.

\bibitem{Shakura and Postnov 2015}  N. Shakura and K. Postnov, A viscous
instability in axially symmetric laminar shear flows, MNRAS 448, 3707-3717,
2015.

\bibitem{Shakura and Postnov II 2015}  N. Shakura and K. Postnov, A
viscousconvective instability in laminar Keplerian thindiscs II. Anelastic
approximation, MNRAS 451, 3995-4004, 2015.

\bibitem{Shakura and Sunyaev 1973}  N. I. Shakura and R. A. Sunyaev, Black
Holes in Binary Systems. Observational Appearance, Astron. Astrophys. 24,
337-355, 1973.

\bibitem{Shapiro and Teukolsky 2004}  Stuart L. Shapiro, Saul A. Teukolsky,
Black Holes, White Dwarfs and Neutron, Wiley-VCH Verlag GmbH \& Co. KGaA,
2004.

\bibitem{Shtemler 2009}  Yuri M. Shtemler, Michael Mond and G\"{u}nther R%
\"{u}diger, Hall equilibrium of thin Keplerian discs embedded in mixed
poloidal and toroidal magnetic fields, Mon. Not. R. Astron. Soc. 394,
1379-1392 (2009).

\bibitem{Shtemler et al. 2010}  Yu. M. Shtemler, M. Mond, G. R\"{u}diger, O.
Regev, and O. M. Umurhan, Non-exponential hydrodynamical growth in
density-stratified thin Keplerian discs, Mon. Not. R. Astron. Soc. 406,
517-528, 2010.

\bibitem{Shtemler 2011}  Yuri M. Shtemler, Michael Mond and Edward Liverts,
Spectral and algebraic instabilities in thin Keplerian discs under poloidal
and toroidal magnetic fields, Mon. Not. R. Astron. Soc. 413, 2957-2977, 2011.

\bibitem{Shtemler 2011b}  Yuri M. Shtemler, Michael Mond, and Edward
Liverts, Regimes of the non-exponential temporal growth in thin Keplerian
discs under toroidally-dominated magnetic fields, arXiv:1109.4719v1
[physics.plasm-ph], 22 Sep 2011.

\bibitem{Shtern at el. 1997}  Vladimir Shtern, Anatoly Borissov, and Fazle
Hussain, Vortex sinks with axial flow: Solution and applications, Phys.
Fluids 9, 2941-2949,1997.

\bibitem{Shore 2007}  Steven N. Shore, Astrophysical Hydrodynamics,
WILEY-VCH Verlag GmbH \& Co KGaA, Weinheim, 2007.

\bibitem{Simon et al. 2001}  M. Simon, A. Dutrey, and S. Guilloteau,
Dynamical Masses of T Tauri Stars and Calibration of Pre-Main-Sequence
Evolution, The Astrophysical Journal, 545:1034-1043, 2001 December 20.

\bibitem{Spruit 2010}  Spruit H. C., Accretion Disks, arXiv:1005.5279v1
[astro-ph.HE] 28 May 2010.

\bibitem{Stahler and Palla 2004}  Steven W. Stahler and Francesco Palla, The
Formation of Stars, WILEY-VCH Verlag GmbH \& Co.KGaA,Weinheim, 2004.

\bibitem{Takakuwa et al. 2012}  Shigehisa Takakuwa, Masao Saito, Jeremy Lim,
Kazuya Saigo, T. K. Sridharan, and Nimesh A. Patel, A Keplerian Circumbinary
Disk around the Protostellar System L1551 NE, The Astrophysical Journal,
754:52 (12pp), 2012 July 20.

\bibitem{Tassoul 2000}  Tassoul Jean-Louis, Stellar Rotation, Cambridge
University Press, 2000.

\bibitem{Tikhonov and Arsenin 1977}  Tikhonov A. N., Arsenin, V. Y.,
Solutions of ill-posed problems, V. H. Winston \& Sons, New York, 1977.

\bibitem{Tominaga and Stathopoulos 2009}  Yoshihide Tominaga, Ted
Stathopoulos, Numerical simulation of dispersion around an isolated cubic
building: Comparison of various types of k-e models, Atmospheric Environment
43 (2009) 3200-3210.

\bibitem{Umurhan and Shaviv  2005}  O. M. Umurhan and G. Shaviv, On the
nature of the hydrodynamic stability of accretion disks, A\&A 432, L31-L34
(2005)

\bibitem{Vasil'eva et al. 1995}  Adelaida B. Vasil'eva, Valentin F. Butuzov,
and Leonid V. Kalachev, The boundary function method for singular
perturbation problems, Society for Industrial and Applied Mathematics, USA,
1995.

\bibitem{Vietri Mario 2008}  Vietri Mario, Foundations of high-energy
astrophysics\textit{, }The University of Chicago Press, Chicago, USA, 2008.

\bibitem{Warsi 1999}  Warsi Z. U. A., Fluid dymnics : theorrttcal and
computational approaches, 2nd ed, CRC Press LLC, USA, 1999.

\bibitem{Wasow 1987}  Wolfgang Richard Wasow, Asymptotic expansions for
ordinary differential equations, Dover Publications, New York, 1987.

\bibitem{White 2006}  Frank M. White, Viscous fluid flow, McGraw-Hill, New
York, 2006

\bibitem{Wilcox 1994}  David C. Wilcox, Turbulence Modeling for CFD, DCW
Industries, USA, California, 1994.

\bibitem{Yen et al. 2014}  Hsi-Wei Yen, Shigehisa Takakuwa, Nagayoshi
Ohashi, Yuri Aikawa, Yusuke Aso, Shin Koyamatsu, Masahiro N. Machida, Kazuya
Saigo, Masao Saito, Kengo Tomida, and Kohji Tomisaka, ALMA Observations of
Infalling Flows toward the Keplerian Disk around the Class I Protostar L1489
IRS, The Astrophysical Journal, 793:1 (20pp), 2014 September 20.

\bibitem{Zhuravlev and Razdoburdin 2014}  V. V. Zhuravlev and D. N.
Razdoburdin, A study of the transient dynamics of perturbations in Keplerian
discs using a variational approach, MNRAS 442, 870-890, 2014.
\end{thebibliography}
\end{document}